\documentclass[preprintnumbers,amsmath,amssymb,floatfix,10pt,prd,onecolumn,superscriptaddress,nofootinbib]{revtex4}
\usepackage{diagbox}
\usepackage{latexsym}
\usepackage{epsfig}
\usepackage{epstopdf,float}
\usepackage{graphicx}
\usepackage{amssymb}
\usepackage{amsmath}
\usepackage{dcolumn}
\usepackage{bm}
\usepackage{color}
\usepackage{comment}
\usepackage[shortlabels]{enumitem}
\usepackage{subfigure}
\usepackage{multirow}
\usepackage{xcolor}
\begin{document}
\title{\bf Shadow Images of Ghosh-Kumar Rotating Black Hole Illuminated By Spherical Light Sources and Thin
Accretion Disks}
\author{Chen-Yu Yang}
\altaffiliation{chenyu\_yang2024@163.com}\affiliation{State Key
Laboratory of Mountain Bridge and Tunnel Engineering, Chongqing
Jiaotong University, Chongqing $400074$,
China}\affiliation{Department of Mechanics, Chongqing Jiaotong
University, Chongqing $400074$, China}
\author{M. Israr Aslam}
\altaffiliation{mrisraraslam@gmail.com}\affiliation{Department of
Mathematics, COMSATS University Islamabad, Lahore Campus,
Lahore-$54000$ Pakistan.}
\author{Xiao-Xiong Zeng}
\altaffiliation{xxzengphysics@163.com}\affiliation{State Key
Laboratory of Mountain Bridge and Tunnel Engineering, Chongqing
Jiaotong University, Chongqing $400074$,
China}\affiliation{Department of Mechanics, Chongqing Jiaotong
University, Chongqing $400074$, China}
\author{Rabia Saleem}
\altaffiliation{rabiasaleem@cuilahore.edu.pk}\affiliation{Department
of Mathematics, COMSATS University Islamabad, Lahore Campus,
Lahore-$54000$ Pakistan.}

\begin{abstract}
This study investigates the astronomical implications of the
Ghosh-Kumar rotating Black Hole (BH), particularly its behaviour on
shadow images, illuminated by celestial light sources and equatorial
thin accretion disks. Our research delineates a crucial correlation
between dynamics of the shadow images and the parameters $a$,~
$q$ and the $\theta_{obs}$, which aptly reflect the influence of the
model parameters on the optical features of shadow images.
Initially, elevated behavior of both $a$ and $q$ transforms the geometry of
the shadow images from perfect circles to an oval shape and
converges them towards the centre of the screen. By imposing the
backward ray-tracing method, we demonstrate the optical appearance
of shadow images of the considering BH spacetime in the celestial
light source. The results demonstrate that the Einstein ring shows a
transition from an axisymmetric closed circle to an arc-like shape
on the screen as well as producing the deformation on the shadow
shape with the modifications of spacetime parameters at the fixed
observational position. Next, we observe that the attributes of
accretion disks along with the relevant parameters on the shadow
images are illuminated by both prograde and retrograde accreting
flow. Our study reveals the process by which the accretion disk
transitions from a disk-like structure to a hat-like shape with the
aid of observational angles. Moreover, with an increase of $q$, the
observed flux of both direct and lensed images of the accretion
disk gradually moves towards the lower zone of the screen.
Furthermore, we present the intensity distribution of the redshift
factors on the screen. Our analysis suggests that the observer can see
both redshift and blueshift factors on the screen at higher
observational angles, while augmenting the values of both $a$ and
$q$, enhancing the effect of redshift on the screen.\\


\end{abstract}
\date{\today}
\maketitle

\section{Introduction}

Black hole, which is regarded as one of the most incomprehensible compact objects
in the universe, have received a huge amount of observational
evidence to present. In $2019$, the Event Horizon Telescope (EHT)
has been instrumental in capturing the images of the super-massive
BH at the heart of the galaxy Messier $87$ (M$87$), based on $1.3$
interferometric observations
\cite{gkref1,gkref2,gkref3,gkref4,gkref5,gkref6}. This is
a substantial evidence of the presence of BHs, and it raises the
compelling prospect of testing General Relativity (GR) in the future
high resolution images of BHs. Moreover, EHT reported that the
configuration of the magnetic field, surrounding the accretion disk,
clarifying that the magnetically seized accretion disk is consistent
with the predictions of general relativistic magnetohydrodynamic
simulations models \cite{gkref7,gkref8}. Recently, EHT released
the first horizon-scale high resolution image of the Sagittarius
(Sgr $A^{\ast}$), which is positioned at the heart of the milky way
and showed that the angular size of the photon ring of the Sgr
$A^{\ast}$ is consistent with the radius of the shadow critical
curve, as calculated in GR within the range of $10\%$ nicely
\cite{gkref9,gkref10,gkref11,gkref12,gkref13,gkref14}. Under the shadow of these
discoveries, it is reasonable to hypothesize that the accretion disk
surrounding the BH could be influencing the observational
characteristics of the BH shadows.

Since observations are susceptible to astronomical
environments around BHs, the intensity to which the current
optical appearance of bright rings is enforced by the photon
rings is still a debatable topic in the scientific community
\cite{gkref15,gkref16,gkref17}. Hence, it is necessary to analyze
the physical properties of astronomical environments
and their contributions to the BH shadows in more detail. As it is well-known in
literature, a supermassive BH surrounding a radiating plasma
results in a luminous accretion disk. The thermal
synchronised electrons are thought to be the primary source of light
in BH image investigation. Moreover, in the case of a high
rotating BH, the emitted radiation can induce a phenomenon,
so-called a relativistic jet \cite{gkref18}. The jet's base is
surrounded by the funnel wall, which can produce not
only potentially triggers substantial amounts of thermal
electromagnetic radiation, but also transforms the jet into a precise
metamorphosis. In this scenario, the jet and accretion disc both
contribute as essential background light sources that are vital to
the BHs horizon-scale imaging procedures. It is widely acknowledged
that in the surroundings of the rotating BHs, there prevails a photon
region, which consists of orbits where photons move at a
continuously constant radial distance, known as bound photon orbits.
This photon region is also known as the photon sphere in the case of static
BHs. When these orbits undergo a radial unstable
perturbations, then the photons on contiguous orbits, that are not
perfectly in the photon zone, may experience multiple half orbits
around the bound orbit before either emerging or falling into the
BH, are the so-called nearly confined bound photon orbits. And the
photons that are close to the bound orbits manifest as an angular
aspect on the observer screen, are almost near the
critical curve, known as the photon ring
\cite{gkref19,rtb30,gkref20}.

The possible observational properties of the BHs shadows and
their related consequences were investigated for a long time. In
$1979$, Luminet considered the radiation from a thin accretion disk
surrounding the BH and obtained simulated images of a BH with thin
accretion disk, which illustrates that the emergence of the BH shadow
depends on the accretion flow and the outer edge of the central
brightness depression \cite{gkref21}. The shadow images for Kerr
spacetime with Keplerian accretion disk have been discussed in detail in Refs.
\cite{gkref22,gkref23}. Based on the general relativistic
ray-tracing code, the authors in \cite{gkref24} investigated the shadow images
of a geometrically thick and infinite thin accretion disk around the
compact objects. Recently, wang et al. \cite{rtb23}
analyzed shadows of Schwarzschild BH surrounded by a Bach-Weyl
ring with the help of the backward ray-tracing procedure. The
Bach-Weyl ring is a concentric thin ring that looks like a Newtonian ring
of constant density, as introduced in \cite{gkref25}. Inspired by
these groundbreaking studies, many other theoretical constraints
on BH shadows and some related geometrical properties are discussed in
the community, including the shadows with
different accretion flow matters and Einstein rings
\cite{gkref26,gkref27,gkref28,gkref29,gkref30,gkref55,gkref56,gkref31,gkref32,gkref33}
etc., based on theoretically well-motivated models. Moreover,
the BH shadow has also been analyzed with some interesting physical
properties in \cite{gkref58,gkref59,gkref60,gkref61,gkref62}. These
heuristics configurations provide us with a realistic description of the
shadow dynamics simulations as well as the influences of accretion
disk on the BH images.

The theory of GR is regarded as a successful theory only for
explaining the gravitational interaction between submillimeter and
solar system scales \cite{gkref34}. However, Planck's length is
generally expected to be replaced by the quantum theory of gravity
with ultraviolet completion \cite{gkref35}. Owing to the absence of
fundamental quantum gravity, extended theories of gravity, as a
phenomenological model representing a classical extension of the GR,
were constructed by complying with the recent observational data and
the data from local tests. Therefore, it is more interesting
to analyze the accreting matter around the BHs in modified
gravities. Hence, this study aims to
deal with this gap in literature. In the fabric of modified theories,
the charged BHs governed by nonlinear electrodynamics (NED) have
recently obtained more significance with the development of
astrophysical observations. This is the reason for the fact that
astrophysical BHs are often carrying strong magnetic fields or immersed
in a plasma medium. Therefore, it is interesting to investigate the
intrinsic properties of BHs described by NED \cite{gkref36}. In
Maxwell's electrodynamics, there exists a singularity at the position
of the point charge and has infinite self energy. To deal with this
issue, Born and Infeld constructed a nonlinear electromagnetic field
\cite{gkref37}. Inspired by this pioneering study, the combination of
GR and Born-Infeld field has been investigated to overcome
the singularity problem of the BHs as well as some other properties
see Refs. \cite{gkref38,gkref39,gkref40}. Moreover, in the action of
effective field theory coming from superstrings, an extended
Born-Infeld action occurs naturally as the leading part
\cite{gkref41,gkref42,gkref43}.

From the astrophysical point of view, the influence of NED become
quite important in super-strongly magnetized compact objects like pulsars and magnetars \cite{gkref44,gkref45}. More importantly,
the first regular BH solution was also investigated as a
spherically symmetric solution to GR source by NED, known as
Bardeen spacetime \cite{gkref46}. Numerous
publications devoted to investigate the intense properties of NED-
sourced BHs (see for example Refs. \cite{gkref47,gkref48,gkref49}).
In this regard, Ghosh and Kumar \cite{rtb1} recently proposed a
magnetically charged rotating BH model (known as GK BH), coupling the
NED to gravity. This BH has a similar type of singularities as
illustrated by Kerr BH \cite{gkref50}. The shadow dynamics cast by
this BH has been discussed in \cite{rtb2}. In addition, this BH
spacetime is closer to Kerr-Newman BH \cite{gkref51} as compared to
newly proposed NED triggered rotating regular BHs
\cite{gkref52,gkref53}. More recently, the author in \cite{rtb6}
considered the GK rotating BH and studied the scattering states of a
massive scalar field and the superradiant amplification
process. With the above motivations, this paper aims to investigate
the optical features of GK rotating BH spacetime with a spherical
background light source along with a thin accretion disk.

The paper will be completed in the following manner. In the upcoming
section, we briefly define the review of GK rotating BH coupled with
NED and then we investigate the effect of spacetime parameters on
the BH shadow images constructed in celestial spheres. In the same
section, we will discuss the simulated images of BH shadow within
the spherical background light source, which are constructed by
imposing the backward ray-tracing method. By considering
geometrically and optically thin accretion disk, in section
\textbf{III}, we will discuss the observed luminosity of the BH
shadow images for values of the associated parameters. Furthermore,
a comprehensive investigation will be conducted on the redshift
factors for both direct and lensed images of the accretion disk in
section \textbf{IV}. The last section is devoted to conclusions and discussions.

\section{CHARGED ROTATING GHOSH-KUMAR BLACK HOLE AND NULL GEODESICS}

\begin{figure}[H]\centering
\subfigure[\tiny][]{\label{alb1}\includegraphics[width=8.5cm,height=8cm]{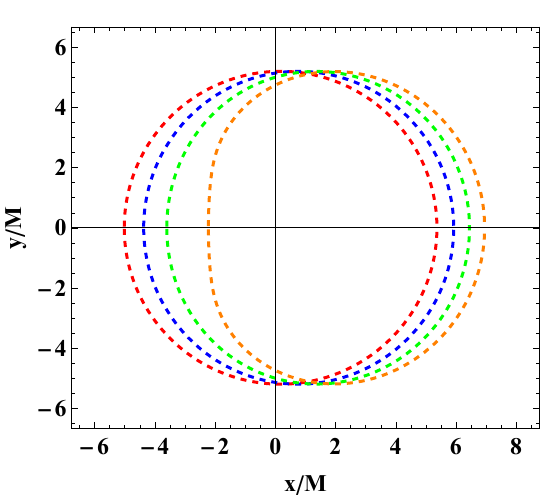}}
\subfigure[\tiny][]{\label{alb2}\includegraphics[width=8.5cm,height=8cm]{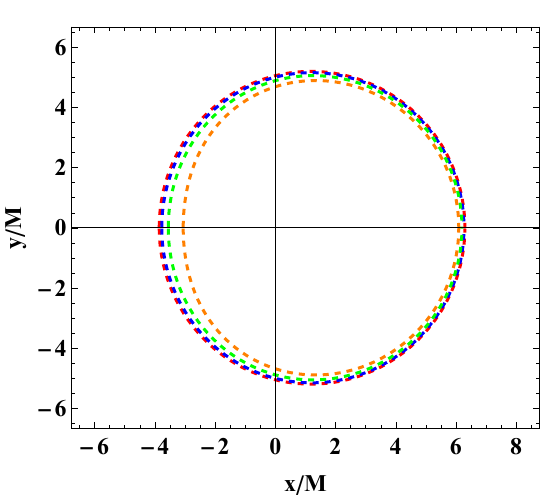}}
\caption{Figure (a) illustrates the circular orbits for different
values of $a=0.1$,~$0.4$,~$0.7$,~$1$, corresponding to red, blue,
green and orange dashed curves, respectively for fixed $q=0.1$.
Whereas Fig. (b) illustrates the circular orbits for different
values of $q=0.1$,~$0.4$,~$0.7$,~$1$, corresponding to red, blue,
green and orange dashed curves, respectively with fixed $a=0.6$.
Both cases are observed for fixed $M=1$ and
$\theta_{obs}=80^{0}$.}\label{shd1}
\end{figure}

The action of GR coupled with NED field is defined as
\cite {rtb3}
\begin{equation}\label{plo1}
S=\int\sqrt{-g}d^4x
\big(\frac{1}{16\pi}\mathcal{R}-\frac{1}{4\pi}\mathcal{L}(F)\big),
\end{equation}
where $\mathcal{R}$ and $g$ represent the Ricci scalar and the
determinant of the metric tensor, respectively. Further, the term
$\mathcal{L}(F)$ represents the lagrangian density of the NED having
function of the form $F=\frac{1}{4}F^{\mu\nu}F_{\mu\nu}$ with
$F_{\mu\nu}=\partial_{\mu}A_{\nu}-\partial_{\nu}A_{\mu}$, where
$A_{\mu}$ is the four-potential. The explicit form of this
Lagrangian density can be defined as \cite {rtb1}
\begin{equation}\label{plo2}
\mathcal{L}=\frac{4M\sqrt{q}F^{\frac{5}{4}}}{q\bigg(\sqrt{2}+2q\sqrt{F}\bigg)^{\frac{3}{2}}},
\end{equation}
in which $M$ and $q$ are BH's mass and magnetic charge, respectively.
Applying Newman-Janis algorithm \cite {rtb4} or its extended
form \cite {rtb5}, one can derive the rotating magnetically charged
GK BH in Boyer-Lindquist coordinate, which can be expressed as \cite
{rtb1,rtb6}
\begin{eqnarray}\nonumber
ds^{2}&=&-\bigg(1-\frac{2Mr^{2}}{\Sigma(r^{2}+q^{2})^{1/2}}\bigg)dt^{2}+\frac{\Sigma}{\Delta}dr^{2}+
\Sigma(r,\theta)d\theta^{2}\\\label{plo3}&-&\frac{4aMr^{2}\sin^{2}\theta}{\Sigma(r^{2}+q^{2})^{1/2}}dtd\phi+
\frac{A(r,\theta)}{\Sigma}\sin^{2}\theta d\phi^{2},
\end{eqnarray}
where
\begin{eqnarray}\nonumber
\Sigma&=&r^{2}+a^{2}\cos^{2}\theta,\\\nonumber
\Delta&=&r^{2}+a^{2}-\frac{2Mr^{2}}{(r^{2}+q^{2})^{1/2}},\\\nonumber
A(r,\theta)&=&(r^{2}+a^{2})^{2}-\Delta a^{2}\sin^{2}\theta,
\end{eqnarray}
in which $a$ is the spin parameter. The above metric recovers the
Kerr BH solution when $q\rightarrow 0$ and the schwarzschild BH
solution when $a=q=0$ \cite {rtb1}. To study the
dynamics of shadows and their associated consequences for rotating BH
in strong field approximation, we need to derive geodesic equations
by considering Hamilton-Jacobi formulation \cite{rtb7}. The
Hamilton-Jacobi equation is given by
\begin{eqnarray}\label{plo4}
\frac{\partial S}{\partial \tau}+H=0, \quad H=
\frac{1}{2}g_{\mu\nu}p^{\mu}p^{\nu}=-\frac{1}{2}m^{2},
\end{eqnarray}
where $\tau$ is the affine parameter, $S$ is the Jacobi action, $H$
is the Hamiltonian, and $p^{\mu}$ is the four-momentum of the
particle and $m$ denotes mass of the particle moving in the BH
space-time. The component $p^{\mu}$ is defined as
\begin{eqnarray}\label{plo5}
p_{\mu}=\frac{\partial S}{\partial
x^{\mu}}=g_{\mu\nu}\frac{dx^{\nu}}{d\tau}.
\end{eqnarray}
Since the Hamiltonian $H$ is independent of $t$ and $\phi$, so there
are two Killing vectors fields such as, $\partial_{t}$ and
$\partial_{\phi}$ as translational and rotational invariance of
time, produce two constants of motion: conserved energy $E=-p_{t}$
and conserved angular momentum $L=p_{\phi}$ about the axis of
symmetry \cite{rtb8}. The components $p_{t}$ and $p_{\phi}$ are
generalized momenta in respective directions. The Jacobi action $S$
of the photon can be separated into the following form
\begin{equation}\label{plo6}
S=\frac{1}{2}m^2\tau-E t+L\phi+D_r(r)+D_{\theta}(\theta),
\end{equation}
where the functions $D_r(r)$ and $D_{\theta}(\theta)$ depend only
on $r$ and $\theta$, respectively yet to be determined. For
photon, $m=0$. Putting Eq. (\ref{plo6}) in Hamilton-Jacobi
equation and then by introducing the Carter constant $\mathcal{Q}$
\cite{rtb9}, one obtains
\begin{eqnarray}\label{plo7}
D_r(r)=\int^{r}\frac{\sqrt{R(r)}}{\Delta}dr, \quad
D_\theta(\theta)=\int^{r}\sqrt{\Theta(\theta)}d\theta,
\end{eqnarray}
with
\begin{eqnarray}\nonumber
R(r)&=&(E(r^{2}+a^{2})-aL)^{2}-\Delta(\mathcal{Q}+(L-aE)^{2}),\\\label{plo8}
\Theta(\theta)&=&\mathcal{Q}+\bigg(a^{2}E^{2}-\frac{L^{2}}{\sin^{2}\theta}\bigg)\cos^{2}\theta,
\end{eqnarray}
where $\Delta$ is defined in Eq. (\ref{plo3}). Then, the variations
of the Jacobi action give rise to the following geodesic equations,
given as
\begin{eqnarray}\nonumber
\Sigma\frac{dt}{d\tau}&=&a(L-aE\sin^{2}\theta)+\frac{r^{2}+a^{2}}{\Delta}(E(r^{2}+a^{2})-aL),\\\nonumber
\Sigma\frac{dr}{d\tau}&=&\pm\sqrt{R(r)},\\\nonumber
\Sigma\frac{d\theta}{d\tau}&=&\pm\sqrt{\Theta(\theta)},\\\label{plo9}
\Sigma\frac{d\phi}{d\tau}&=&\frac{L^{2}}{\sin^{2}\theta}-aE+\frac{a}{\Delta}(E(r^{2}+a^{2})-aL).
\end{eqnarray}
The sign $\pm$ corresponds to the positive and negative momentum,
Respectively, i.e., the direction of moving particle. Since we are
interested in analyzing the boundary of the shadow cast by the
rotating GK BH, so we consider the geodesic equation which can be
written as
\begin{equation}\label{plo10}
\frac{1}{2}(\Sigma\frac{dr}{d\tau})^{2}+V(r)=0,
\end{equation}
where $V(r)$ is the effective potential that can be defined as $V(r)=-R(r)/2r^{4}$ and in the equatorial plane ($\theta=\pi/2$) . The maximum
value of the $V(r)$ corresponds to the position of the unstable
photon orbit. The BH shadow silhouette is derived from the specific
orbit in the radial equation, defined by $r=r_{ph}$, which satisfies
the underlying conditions \cite{rtb11}
\begin{eqnarray}\label{plo12}
V(r)|_{r=r_{ph}}=0, \quad \frac{\partial V(r)}{\partial
r}|_{r=r_{ph}}=0.
\end{eqnarray}
The motion of a photon is determined by two impact parameters
\begin{eqnarray}\label{plo13}
\xi=\frac{L}{E}, \quad \quad \eta = \frac{\mathcal{Q}}{E^2},
\end{eqnarray}
which parameterize the null geodesics. Solving Eq.
(\ref{plo12}), one can obtain the following expressions of the
critical values of impact parameters ($\xi_{cr},~\eta_{cr}$) for the
unstable photon orbits as
\begin{eqnarray}\label{plo14}
\xi_{cr}&=&\frac{(a^{2}+r^{2}_{ph})\Delta'(r_{ph})-4r_{ph}\Delta(r_{ph})}{a\Delta'(r_{ph})},\\\nonumber
\eta_{cr}&=&(r^{2}_{ph}(-16\Delta(r_{ph})^{2}-r^{2}_{ph}\Delta'(r_{ph})^{2}+8\Delta(r_{ph})(2a^{2}\\\label{plo15}&+&
r_{ph}\Delta'(r_{ph}))))(a^{2}\Delta'(r_{ph})^{2})^{-1},
\end{eqnarray}
here ($'$) denotes the derivative with respect to $r$. Now, we are
going to analyze the significant features of the free-moving matter
around BH spacetime geometry. In general, the photons emitted by a
light source are deflected when they pass near a BH due to strong
gravitational lensing. Some of them, after being deflected by the
BH, can reach a distant observer, while others fall directly into
the BH. The photons that cannot escape from the BH shadow are in the
observer's sky. So, we introduce the celestial coordinates,
connected to the actual astronomical measurements that span a two
dimensional plane, written as \cite{rtb12,rtb13}
\begin{eqnarray}\label{plo16}
X &=& \lim_{r_{obs}\rightarrow \infty}\bigg(-r^2_{obs}
\sin\theta_{obs}\left(\frac{d\phi}{dr}\right)_{(r_{obs},\theta_{obs})}\bigg),\\\label{plo17}
Y &=& \lim_{r_{obs}\rightarrow
\infty}\bigg(r^2_{obs}\left(\frac{d\theta}{dr}\right)_{(r_{obs},\theta_{obs})}\bigg),
\end{eqnarray}
here $r_{obs}$ is the radial coordinate position of the observer, lie far
away from the BH, and $\theta_{obs}$ is the inclination angle of the
observer. Now Eqs. (\ref{plo16}) and (\ref{plo17}) can be
further simplified in terms of impact parameters as \cite{rtb13}
\begin{eqnarray}\label{plo18}
X &=&-\xi_{cr}\csc\theta_{obs},\\\label{plo19} Y
&=&\pm\sqrt{\eta_{cr}+a^{2}\cos^{2}\theta_{obs}-\xi^{2}_{cr}\cot^{2}\theta_{obs}}.
\end{eqnarray}
If the observer is located on the equatorial plane, i.e.,
$\theta_{obs}\rightarrow\pi/2$, the above equations can be
simplified as
\begin{eqnarray}\label{plo20}
X &=&-\xi_{cr},\\\label{plo21} Y &=&\pm\sqrt{\eta_{cr}}.
\end{eqnarray}
To analyze the shape and size of the shadow for an equatorial
observer in the strong field region, we plot the shadow images of
rotating GK BH for various values of $a$ and $q$ in Fig.
\textbf{\ref{shd1}}. The left panel shows the circular orbit for
different values of spin parameter $a$. It is observed that with
the aid of $a$, the circular orbits are shifted in the positive
$x$-direction. For smaller values of $a$, the shadows
images can be identified as perfect circles. However, when $a=1$, the
shadow images move rightwards with an elongation in the shadow and
exhibit a clear possible flatness on the left side of the shadow,
see Fig. \textbf{\ref{shd1}} (a). In Fig. \textbf{\ref{shd1}} (b),
very small differences are measured in the shadows with the
variation of NED parameter $q$. The image size reduced with
increasing $q$ and the shadow images can be identified as
perfect circles when $q$ has lesser values. With an increase in the value of
$q$, the shadow images are slightly moved towards the positive
$x$-direction and negligible flatness is being observed on the left side
of the shadow images when $q\rightarrow 1$. This deforms the
perfect circles into oval geometry. These effects
and differences of images will be discussed in the upcoming
sections, which may serve as a criterion to distinguish GK rotating
BHs from other spacetimes.

\subsection{Shadow Images Under Spherical Background Light Source}

\begin{figure*}
\begin{center}
\subfigure[\tiny][]{\label{a1}\includegraphics[width=4.1cm,height=4.2cm]{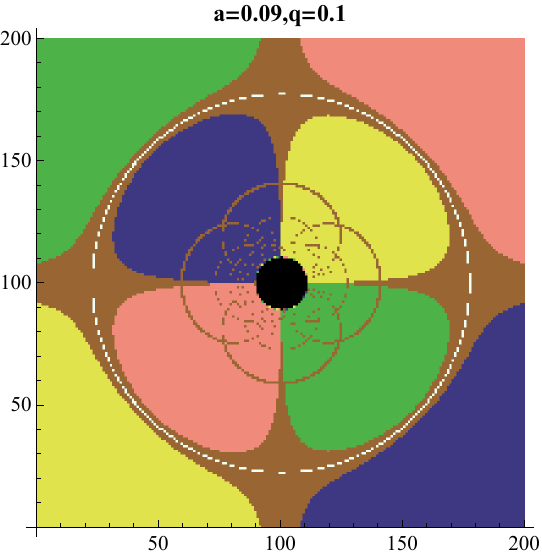}}
\subfigure[\tiny][]{\label{b1}\includegraphics[width=4.1cm,height=4.2cm]{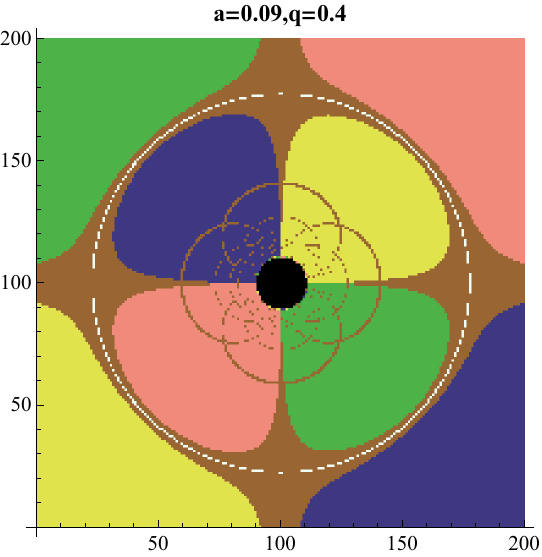}}
\subfigure[\tiny][]{\label{c1}\includegraphics[width=4.1cm,height=4.2cm]{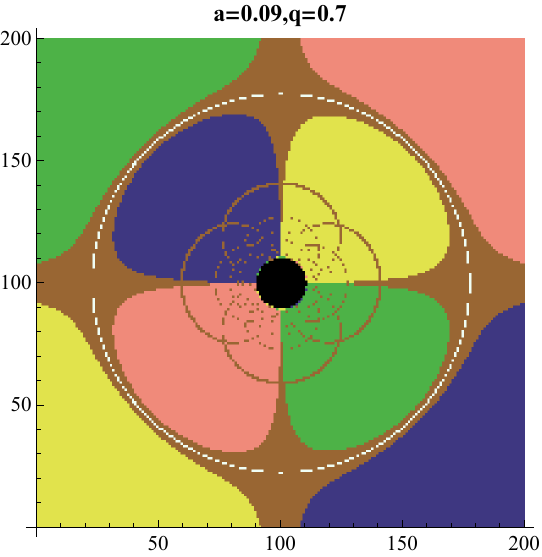}}
\subfigure[\tiny][]{\label{a1}\includegraphics[width=4.1cm,height=4.2cm]{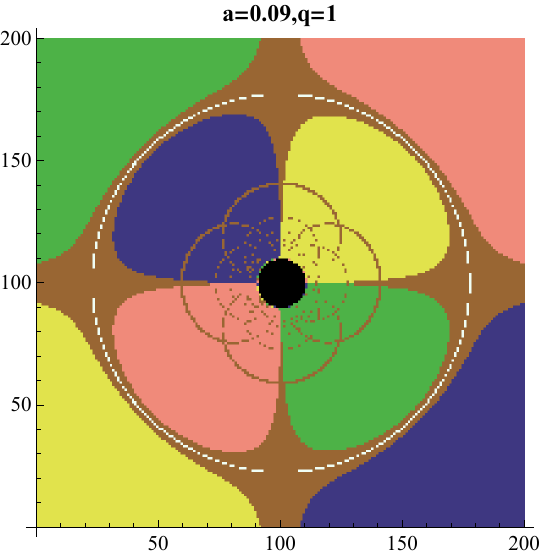}}
\caption{The shadow images cast by rotating GK BH for different
values of $q$. Here, we fixed $M=1$, $a=0.09$, $r_{obs}=500M$ and
the inclination angle of the observer is
$\theta_{obs}=80^{0}$.}\label{shd2}
\end{center}
\end{figure*}

\begin{figure*}
\begin{center}
\subfigure[\tiny][]{\label{a1}\includegraphics[width=4.1cm,height=4.2cm]{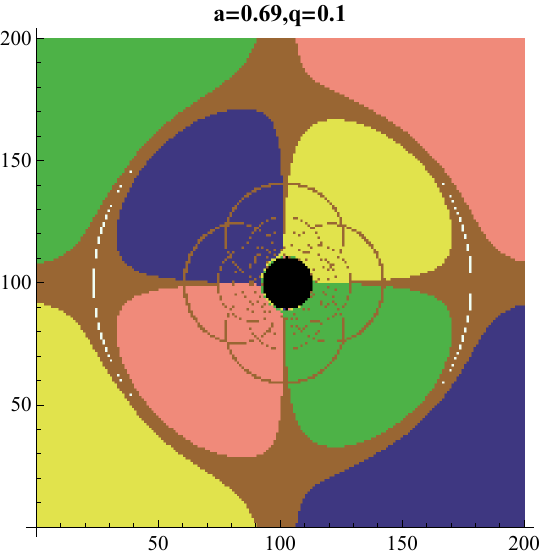}}
\subfigure[\tiny][]{\label{b1}\includegraphics[width=4.1cm,height=4.2cm]{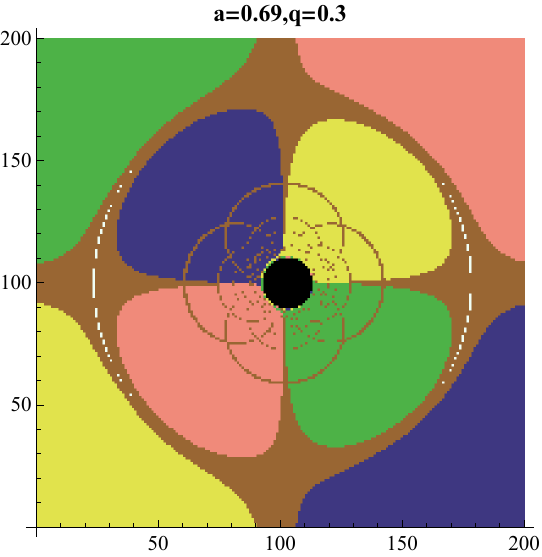}}
\subfigure[\tiny][]{\label{c1}\includegraphics[width=4.1cm,height=4.2cm]{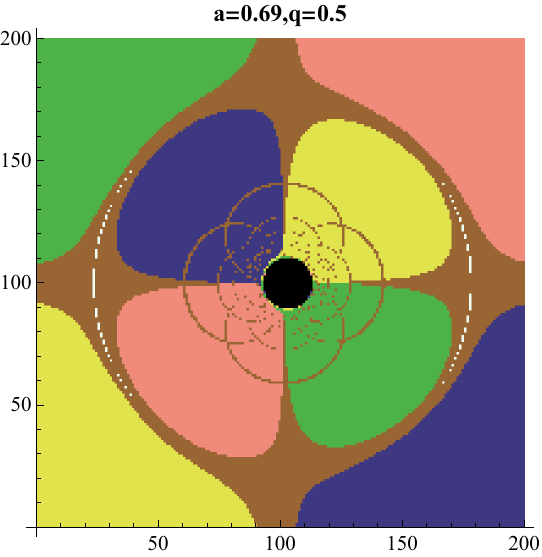}}
\subfigure[\tiny][]{\label{a1}\includegraphics[width=4.1cm,height=4.2cm]{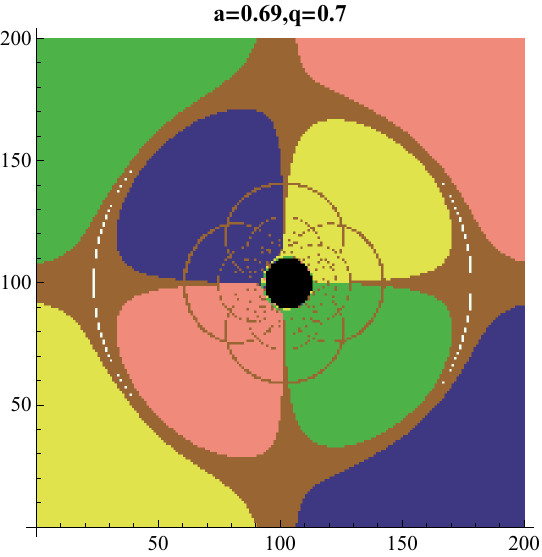}}
\caption{The shadow images cast by rotating GK BH for different values of $q$. Here, we fixed $M=1$, $a=0.69$,
$r_{obs}=500M$ and the inclination angle of the observer is
$\theta_{obs}=80^{0}$.}\label{shd3}
\end{center}
\end{figure*}

\begin{figure*}
\begin{center}
\subfigure[\tiny][]{\label{a1}\includegraphics[width=4.1cm,height=4.2cm]{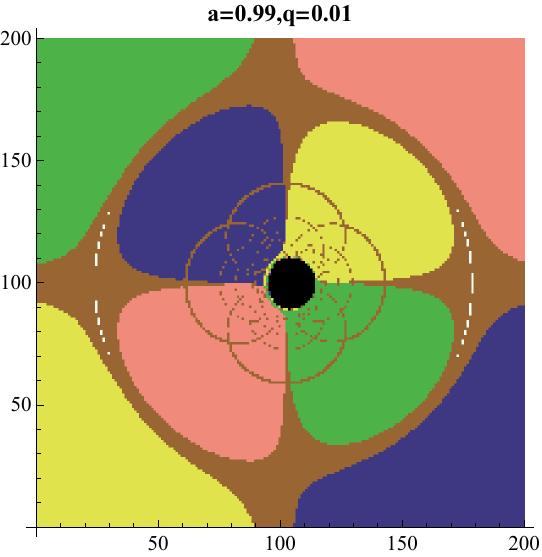}}
\subfigure[\tiny][]{\label{b1}\includegraphics[width=4.1cm,height=4.2cm]{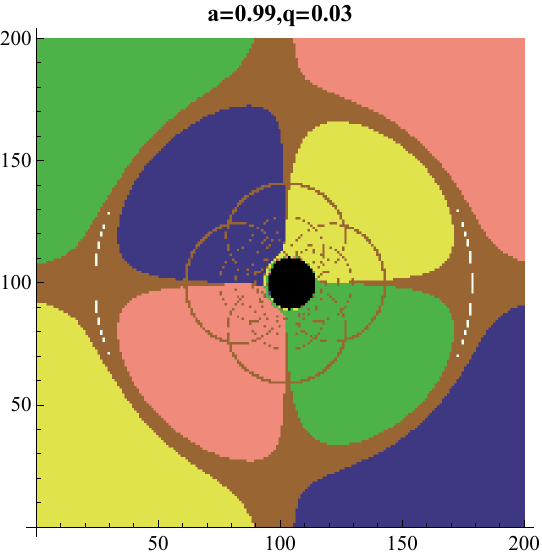}}
\subfigure[\tiny][]{\label{c1}\includegraphics[width=4.1cm,height=4.2cm]{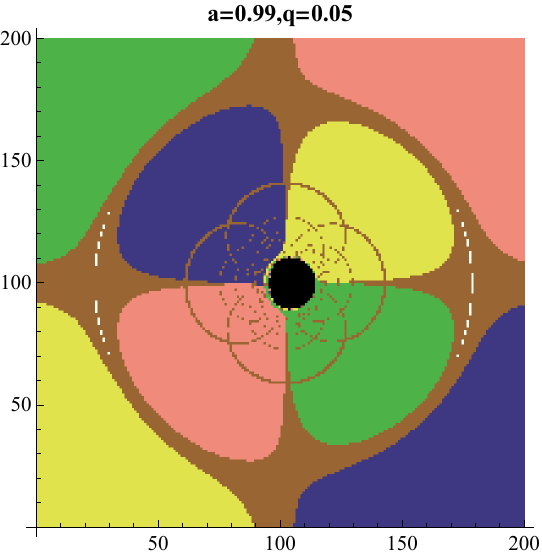}}
\subfigure[\tiny][]{\label{a1}\includegraphics[width=4.1cm,height=4.2cm]{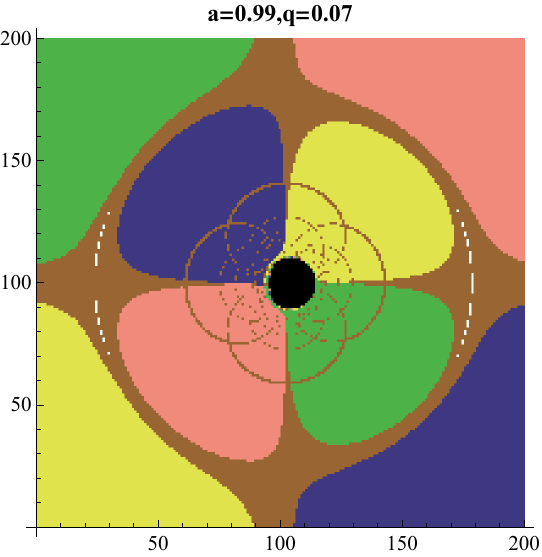}}
\caption{The shadow images cast by rotating GK BH
for different values of $q$. Here, we fixed $M=1$, $a=0.99$,
$r_{obs}=500M$ and the inclination angle of the observer is
$\theta_{obs}=80^{0}$.}\label{shd4}
\end{center}
\end{figure*}

Now we discuss the optical appearance of BH shadow images with the
help of a well-known numerical backward ray-tracing method which has
been widely used in literature \cite{rtb14,rtb15,rtb16,rtb17,rtb18,rtb19,rtb20,rtb21,rtb22,rtb23,rtb24}.
In the present study, we will provide the essential steps of this
method, which may be fruitful for the readers. In this setup, one can
assume that the light rays evolve from the observer reverse in time. After that, one can define the position of each pixel in the
final image by solving the null geodesic equations
numerically. Here, the optical appearance of BH shadow in
the observer's screen is suppressed by the pixels connected to the
photons that directly fall into the BH. We assume that an
observer is located at ($r_{obs},~\theta_{obs}$) in the coordinated
($t,~r,~\theta,~\phi$). In this scenario, one can expand the
observer basis
($e_{\widehat{t}},~e_{\widehat{r}},~e_{\widehat{\theta}},~e_{\widehat{\phi}}$)
as a form on the coordinate basis
\cite{rtb14,rtb15,rtb16,rtb17,rtb18,rtb19,rtb20,rtb21,rtb22,rtb23,rtb24}
\begin{equation}\label{plo22}
e_{\widehat{\mu}}\equiv e^{\nu}_{\widehat{\mu}}\partial_{\nu},
\end{equation}
in which the matrix $e^{\nu}_{\widehat{\mu}}$ satisfies
$g_{\mu\nu}e^{\mu}_{\widehat{\xi}}e^{\nu}_{\widehat{\zeta}}=\varpi_{\widehat{\xi}\widehat{\zeta}}$,
where $\varpi_{\widehat{\xi}\widehat{\zeta}}$ is the Minkowski
spacetime. For spacetime (\ref{plo3}), it is convenient to choose a decomposition
\begin{eqnarray}
\label{plo23} e^{\nu}_{\widehat{\mu}}=\left(\begin{array}{cccc}
\sigma&0&0&\rho\\
0&\mathcal{A}^r&0&0\\
0&0&\mathcal{A}^{\theta}&0\\
0&0&0&\mathcal{A}^{\phi}
\end{array}\right),
\end{eqnarray}
here $\sigma$, $\rho$, $\mathcal{A}^r$, $\mathcal{A}^{\theta}$, and
$\mathcal{A}^{\phi}$ are real coefficients. Further, Eq.
(\ref{plo23}) is related to zero angular momentum observer in
relation to spatial infinity. Based on the Minkowski normalization
$e_{\widehat{\mu}}e^{\widehat{\nu}}=\delta^{\widehat{\nu}}_{\widehat{\mu}}$,
one can obtain the following relations
\begin{eqnarray}\nonumber
&&\mathcal{A}^r=\frac{1}{\sqrt{g_{rr}}},\;\;\;\;\;\;\;\;\;\;\;
\mathcal{A}^{\theta}=\frac{1}{\sqrt{g_{\theta\theta}}},\;\;\;\;\;\;\;\;\;\;\;
\mathcal{A}^{\phi}=\frac{1}{\sqrt{g_{\phi\phi}}},\;\;\;\;\;\;\;\;\;\;\;\\\label{plo24}&&
\sigma=\sqrt{\frac{g_{\phi\phi}}{g^2_{t\phi}-g_{tt}g_{\phi\phi}}},\;\;\;\;\;\;\;\;\;\;\;
\rho=-\frac{g_{t\phi}}{g_{\phi\phi}}\sqrt{\frac{g_{\phi\phi}}{g^2_{t\phi}-g_{tt}g_{\phi\phi}}}
\end{eqnarray}
From Eq. (\ref{plo22}), the locally measured $4$-momentum of a
photon $p^{\widehat{\mu}}$ can be defined as
\begin{eqnarray}\label{plo25}
p^{\widehat{t}}=-p_{\nu}e^{\nu}_{\widehat{t}},
\;\;\;\;\;\;\;\;\;\;\;p^{\widehat{i}}= p_{\nu}e^{\nu}_{\widehat{i}},
\end{eqnarray}
where ($\widehat{i}=r,~\theta,~\phi$). Hence, the locally measured
$4$-momentum of coupled photons $p^{\widehat{\mu}}$ can be
redefined as
\begin{eqnarray}\nonumber
&&p^{\widehat{t}}=\sigma E-\rho
L,\;\;\;\;\;\;p^{\widehat{\phi}}=\frac{L}{\sqrt{g_{\phi\phi}}},\\\label{plo26}&&
p^{\widehat{\theta}}=\frac{p_{\theta}}{\sqrt{g_{\theta\theta}}},
\;\;\;\;\;\;\; p^{\widehat{r}}=\frac{p_{r}}{\sqrt{g_{rr}}},
\end{eqnarray}
where $p_{\theta}$ and $p_{r}$ are the components of momentum of the
photon $p_{\theta}=g_{\theta\theta}\frac{d\theta}{d\tau}$ and
$p_{r}=g_{rr}\frac{dr}{d\tau}$, respectively. The three vector
$\overrightarrow{p}$ is the linear momentum of photons having
components
($p^{\widehat{r}},~p^{\widehat{\theta}},~p^{\widehat{\phi}}$) in the
orthonormal basis
($e_{\widehat{r}},~e_{\widehat{\theta}},~e_{\widehat{\phi}}$) as
\cite{rtb25}
\begin{equation}\label{plo27}
\overrightarrow{p}=p^{\widehat{r}}e_{\widehat{r}}+p^{\widehat{\theta}}e_{\widehat{\theta}}+p^{\widehat{\phi}}e_{\widehat{\phi}}.
\end{equation}
Based on the geometry of the photon's observables, we have
\cite{rtb25}
\begin{eqnarray}\nonumber
p^{\widehat{r}}&=&|\overrightarrow{p}|\cos\mathbb{A}\cos\mathbb{B},\\\nonumber
p^{\widehat{\theta}}&=&|\overrightarrow{p}|\sin\mathbb{A},\\\label{plo28}
p^{\widehat{\phi}}&=&|\overrightarrow{p}|\cos\mathbb{A}\sin\mathbb{B},
\end{eqnarray}
where ($\mathbb{A},~\mathbb{B}$) are the angular coordinates of a
point in the local observer's sky, interpreting the orientation of the
associated photon and develop its initial conditions. Closely
followed by similar steps
\cite{rtb14,rtb15,rtb16,rtb17,rtb18,rtb19,rtb20,rtb21,rtb22,rtb23,rtb24,rtb25},
one can obtain the celestial coordinates ($X,~Y$) for the pixel
corresponding to light ray which is defined as
\begin{eqnarray}\nonumber
X&=&-r_{obs}\tan\mathbb{B}=-r_{obs}\frac{p^{\widehat{\phi}}}{p^{\widehat{r}}},\\\nonumber
Y&=&r_{obs}\frac{\tan\mathbb{A}}{\cos\mathbb{B}}=r_{obs}\frac{p^{\widehat{\theta}}}{p^{\widehat{r}}}\label{plo29}.
\end{eqnarray}
In this scenario, for the spacetime (\ref{plo3}), one can define the
observational position of the image produced due to the photon in
observer's sky as
\begin{eqnarray}\nonumber
X&=&-r_{obs}\frac{p^{\widehat{\phi}}}{p^{\widehat{r}}}=-r_{obs}\frac{\Delta\sqrt{g_{rr}}L}{\sqrt{g_{\phi\phi
R(r_{obs})}}},\\\label{plo30}
Y&=&r_{obs}\frac{p^{\widehat{\theta}}}{p^{\widehat{r}}}=r_{obs}\frac{\Delta\sqrt{g_{rr}}p_{\theta}}{\sqrt{g_{\theta\theta
R(r_{obs})}}},
\end{eqnarray}
where $r=r_{obs}$ and $\theta=\theta_{obs}$. Previously, we
defined the celestial coordinates in Eqs. (\ref{plo16}) and
(\ref{plo17}), these results are employed only when a real observer
lies far away from the BH such as $r_{obs}\rightarrow\infty$. In
this way, we have shown the optical appearance of shadow images
of the considering BH spacetime in the celestial sphere with the help of
backward ray-tracing method in Figs. \textbf{\ref{shd2}}-\textbf{\ref{shd4}} for different values of $a$
and $q$. Similarly, as done in Refs.
\cite{rtb14,rtb15,rtb16,rtb17,rtb18,rtb19,rtb20,rtb21,rtb22,rtb23,rtb24},
for the comprehensive physical interpretation of shadow images, one can
classify the celestial sphere into four different quadrants mark
with colours (\textcolor{pink}{$\spadesuit$},
\textcolor{blue}{$\spadesuit$}, \textcolor{yellow}{$\spadesuit$},
\textcolor{green}{$\spadesuit$}). The grid of both longitude and
latitude lines is denoted by brown lines positioned by $10^{0}$.
The dark area in the center of each panel shows the BH, where
photons directly fall into the BH. In each panel, outside the
``D'' shape petals, there is a white circular curve which could
provide a direct interpretation of Einstein ring. In Fig.
\textbf{\ref{shd2}}, we interpret the shadow casted by rotating GK
BH for different values of parameters $q$ with fixed $a=0.09$.
Here we see that from left to right as the values of
parameter $q$ grow the shadow’s shape is a perfect circle, which is
similar to those of the usual static BHs. However, with the aid
of $q$, from left to right the size of the shadow
decreases negligibly. Importantly, when $q=1$, the Einstein ring is vertically
divided into two equal parts, which shows the effect of NED on
the shadow of the BH. Similarly, when we increase the values of spin
parameter $a=0.69$ and parameter $q=0.1,~0.3,~0.5,~0.7$ varies from
left to right in Fig. \textbf{\ref{shd3}}, we notice that the
Einstein ring shows the arcs like shapes in the middle of the
screen. Moreover, with increasing $q$, from left to right
the shadow shape is a perfect circle and remains the same in all
cases. When we further increase the value of $a=0.99$ and choose the
smaller values of $q$ as compared to previous cases such that,
$q=0.01,~0.03,~0.05,~0.07$ corresponds to Fig. \textbf{\ref{shd4}}
(a-d), again the Einstein ring shows the arcs like shapes in the
middle of the screen. But these arcs are smaller as compared to
the previous one. Again, the size of the shadow is the same and the
white light arc is negligibly dispersed with the aid of $q$. In
all cases, we observed a common phenomenon which is the dark area as
well as ``D'' shape petals remain the same. These distinct features
in the shadow can be attributed to the effect of the spin and NED
parameters i.e., $a$ and $q$ on the spacetime structures. These
results may differentiate the rotating GK BH from other spacetime
structures.

\section{Dynamics of Black Hole Shadow Within A Thin Disk Accreting Flow}

The analysis of accreting disk around BHs provides a crucial
entrance for deepening our comprehension of the vital physical
mechanisms governing these intricate systems. This section is
devoted to advancing our comprehension through an investigation into
the observable attributes of accretion disks within the paradigm of
considering spacetime. We hope this analysis elucidates the
distinctive impact on the observable features of accretion disk
structures that arise from modifications to the geometric
properties of spacetime within the rotating GK BH paradigm. It is
fascinating to analyze numerous specific aspects that we take into
account in this accretion model. Since we assume that the accreting
matter is electrically free neutral plasma and moves along an
equatorial timelike geodesic. Moreover, in the examination of the BH
accretion disk, the radii of the BH provide significant factors
about the shadow of the BH. The inner stable circular orbit (ISCO)
are related to the inner boundary of the accretion disk around the
BH and their radii can be used to compute the energy emission
efficiency, which is a measure of how quickly energy from the rest
mass turns into radiation. The mechanism of these particles during
this process has been analyzed with astrophysical observations
\cite{rtb32}. The ISCO indicates a demarcation line,
outside this the emission spectrum of the accretion disk will
maintain stable circular orbits, while inside the boundary they
will undergo critical plunging orbits. Now, we would like to
define the circular orbits of the particles that lie within the range
$r\geq r_{ISCO}$, which should satisfy the condition, defined in Eq.
(\ref{plo12}). However, for the range $r< r_{ISCO}$, one can define
the radial motion $U_{r}$ as given below
\begin{equation}\label{plo38}
U_{r}=-\sqrt{-\frac{V(r)}{g_{rr}}},
\end{equation}
where the (``$-$'') sign in front of the square root means the inward
motion of particles. Notably, in our examination when tracing the
associated light ray backwards, it may cross the equatorial plane
many times, having different radii of intersections. Letting
$r_{n}(X, Y)$ represents the radius at which a ray intersects the
equatorial plane for the n$^{th}$ time on its backward journey from
screen position $(X, Y)$. Moreover, the function $r_{n}(X, Y)$ is
so-called the radiative transport, which is governed by the
Boltzmann equation for photons. This function generates the
shape of the n$^{th}$ image of the disk. For example, when $n=1$, it
corresponds to \textit{direct} image and $n=2$ corresponds to
\textit{lensed} image, which is produced on the screen. Importantly,
the radiative transfer function closely depends on the observational
angle $\theta_{obs}$ and algorithms used here typically form images,
where each pixel on the image corresponds to a wave vector.

In this scenario, we still consider the screen in the frame of zero
angular momentum and the intensity of the accretion disk naturally
change due to the emission and absorption of light rays when they
reaches the accretion disk. For the sake of convenience, we assume
that the scattering is negligible and the equation which is used to
determined the change of intensity can be defined as the
following familiar form \cite{rtb26,rtb27,rtb28,rtb29,gkref57}
\begin{equation}\label{plo31}
\frac{d\mathcal{I}}{d\tau}=\mathcal{J}-\mathcal{C}\mathcal{I},
\end{equation}
where $\mathcal{I}=I_{\nu}/\nu^{3}$, $\mathcal{J}=J_{\nu}/\nu^{2}$
and $\mathcal{C}=\alpha_{\nu}\nu$. Further, the components $I_{\nu}$
(cgs units: erg cm$^{-2}$ s$^{-1}$ sr$^{-1}$ Hz$^{-1}$), $J_{\nu},$
(cgs units: erg cm$^{-3}$ s$^{-1}$ sr$^{-1}$ Hz$^{-1}$),
$\alpha_{\nu}$ (cgs units: erg cm$^{-1}$) denotes specific
intensity, emissivity per unit volume and absorption coefficient
with frequency $\nu$, respectively. Notice that, when the beams of
light propagating in different directions at a single point in
space, both $J_{\nu},$ and $\alpha_{\nu}$ approaches to zero and
hence the quantities $I_{\nu}/\nu^{3}$ is conserved along the
geodesic.

Moreover, we assume that spinning spacetime is stationary,
axisymmetric, and reflection-symmetric about the equatorial plane.
Since the accretion disk is geometrically thin, the
coefficient of emissivity and absorption remains constant when the
light rays cross it. Integrating Eq. (\ref{plo31}), one can obtain
the final form of intensity on the observer's screen, given as below
(the readers can see the complete derivation in Ref. \cite{rtb28})
\begin{equation}\label{plo32}
I_{\nu_{obs}}=\sum_{n=1}^{m}
\bigg(\frac{\nu_{obs}}{\nu_{n}}\bigg)^{3}\frac{J_{n}}{\Phi_{n}-1}\bigg(\frac{1-e^{-\alpha_{n}\mathcal{F}_{n}}}{\mathcal{F}_{n}}\bigg),
\end{equation}
where $\nu_{obs}$ indicates the observed frequency by the observer,
$\nu_{n}$ represents the frequency observed by the local static
frames co-moving with the emission profiles and $m$ is the maximum
number of intersections on the screen position $(X, Y)$. Further, we
consider the class of frames $\{\mathbb{F}_n\}$, where $n=1...m(X,
Y)$ is the maximum number of times that the light ray passes the
equatorial plane, and the subscript $n$ represents the
corresponding measurements in the static local frames
$\mathbb{F}_n$. The component $\Phi_{n}$ is the optical depth of
photons change accordingly when they are emitted at $N$ defined as
\begin{equation}\label{plo33}
\Phi_{N}=
\begin{cases}
\exp\big(\sum_{n=1}^{N}\alpha_{n}\mathcal{F}_{n}\big) \quad \quad
\text{if}\,\,\, \ N\geq 1,\\
1 \quad \quad \quad \quad \quad \quad \quad \quad \quad \,\,
\text{if}\,\,\, \ N= 0,
\end{cases}
\end{equation}
with $\mathcal{F}_{n}=\nu_{n}\Delta\tau_{n}$ a fudge factor to
account for the effects of geometrical thickness, further related
with to artificially enhances the flux in the photon rings.
Moreover, it is a highly defensible fudge, allowing the equatorial
approximation, leading to the enhancement of its simplicity and speed
\cite{rtb30}. In the fudge factor, the quantity of the component
$\Delta\tau_{n}$ is the change in the affine parameter when the
particular ray passing through the accretion disk medium at
$\mathbb{F}_n$. When we consider the optically
thin accretion disk means that we safely ignore the absorption effect, then
Eq. (\ref{plo32}) reduce to the following form
\begin{equation}\label{plo34}
I_{\nu_{obs}}=\sum_{n=1}^{m}\mathcal{F}_{n}\textbf{g}_{n}^{3}(r_{n},~X,
Y)J_{n},
\end{equation}
where $\textbf{g}_{n}$ is the redshift factor, which is
$\textbf{g}_{n}=\nu_{obs}/\nu_{n}$. Since, the emission spectrum of
particles can be categorized into two different regions based on
the ISCO, the redshift factor of the accretion disk within and
beyond the ISCO display perceptible differences. Firstly, particles lying beyond the ISCO move along a circular orbit having angular velocity $\Gamma$, associated with the redshift factor can
be defined as
\begin{equation}\label{plo39}
\textbf{g}_{cir}=\frac{\widehat{E}}{(1-\Gamma_{n}\textbf{b})\chi},
\quad \quad r_{n}\geq r_{ISCO},
\end{equation}
where
\begin{equation}\label{plo40}
\Gamma_{n}(r)=\frac{U^{\phi}}{U^{t}}|_{r=r_{n}}.
\end{equation}
In Eq. (\ref{plo39}) the term $\textbf{b}$ is the impact parameter,
defining the ratio of energy and angular momentum of photon along
null geodesics, $\chi$ is angular velocity function and
$\widehat{E}$ is the ratio of observed energy on the image plane to
the conserved energy along null geodesics. Mathematically, these
components can be defined as
\begin{eqnarray}\label{plo41}
\textbf{b}=\frac{L_{n}}{E_{n}}, \quad
\widehat{E}=\frac{E_{n}}{E_{n}^{o}}=\sigma+\widehat{E}\rho, \quad
\chi=\sqrt{\frac{-1}{g_{tt}+2g_{t\phi}\Gamma_{n}+g_{\phi\phi}\Gamma_{n}^{2}}}|_{r=r_{n}},
\end{eqnarray}
where $E_{n}^{o}$ is the observed energy of the photon. Importantly,
when $r_{obs}\rightarrow \infty$, then $\widehat{E}\rightarrow 0$
for the asymptotically flat spacetimes. Now, the particles are
moving along the critical plunge orbit, lie inside the ISCO, and then
the redshift factor has the following form
\begin{equation}\label{plo42}
\textbf{g}_{plu}=-\frac{\widehat{E}}{U_{r}p_{r}/E_{n}+E_{i}(g^{tt}-\textbf{b}g^{t\phi})
+L_{i}(\textbf{b}g^{\phi\phi}-g^{t\phi})}|_{r=r_{n}}, \quad \quad
r_{n}< r_{ISCO},
\end{equation}
where $E_{i}$ and $L_{i}$ are the energy and the angular momentum of
the photon on ISCO, respectively. A similar treatment was used to
analyze the images of Kerr BHs in Refs. \cite{rtb30,gkref54}.
Moreover in \cite{rtb28,rtb31,rtb32}, authors employed Eq.
(\ref{plo34}) and analyzed different dynamics of simulated images
of Kerr geometry surrounded by an optically thin accretion disk, having
central brightness depression and a narrow bright photon ring.
Specifically, considering the ratio of the observation wavelength (1.3
mm), which is consistent with the images of M$87$ and Sgr A$^{\ast}$
at $230$ GHz, we define the emissivity to be a second-order
polynomial in log-space as
\begin{equation}\label{plo35}
\log[J]=\big(\rho_{1}k^{2}+\rho_{2}k\big),
\end{equation}
with $k=\log(\frac{r}{r_{+}})$. To interpret
$230$ GHz images in this paper, we set $\rho_{1}=-1/2$ and
$\rho_{2}=-2$ for better visual appearance \cite{rtb28,rtb32}. In
addition, the roots of $\Delta$ correspond to the horizons of the
BH, and we choose here the largest root ($r_{+}$) for the value of
the event horizon of the BH. Since the emission profile is
isotropic and axisymmetric and decaying rapidly with respect to the
horizon $r_{+}$. In \cite{rtb32}, authors worked with different
choices of $\mathbb{F}_n$, we normalize all the fudge factors
$\mathbb{F}_n$ to $1$ \cite{rtb28}. Because we mainly focused on
discussing the influence of magnetic fields on the emission profiles,
as the values of $\mathbb{F}_n$ vary the optical
appearance of the photon ring negligibly, which has very limited influence on
overall image. In this scenario, we use Eq. (\ref{plo34}) and plot
the corresponding emission profiles for different choices of the model
parameters.

\subsection{Optical Signatures of the Black Hole Shadow Images}

\begin{figure*}
\begin{center}
\subfigure[\tiny][$~q=0.09,~\theta_{obs}=0.01^{o}$]{\label{a101}\includegraphics[width=5.9cm,height=5.8cm]{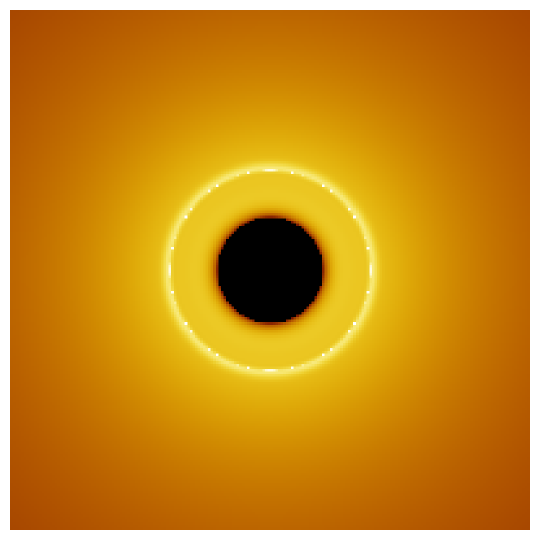}}
\subfigure[\tiny][$~q=0.09,~\theta_{obs}=17^{o}$]{\label{b102}\includegraphics[width=5.9cm,height=5.8cm]{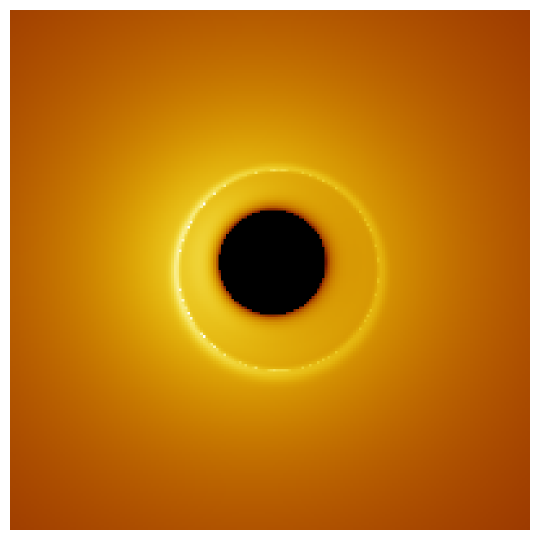}}
\subfigure[\tiny][$~q=0.09,~\theta_{obs}=80^{o}$]{\label{c103}\includegraphics[width=5.9cm,height=5.8cm]{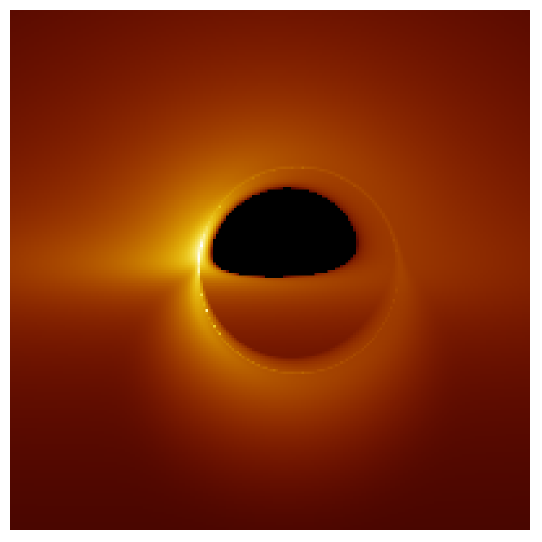}}
\subfigure[\tiny][$~q=0.39,~\theta_{obs}=0.01^{o}$]{\label{a104}\includegraphics[width=5.9cm,height=5.8cm]{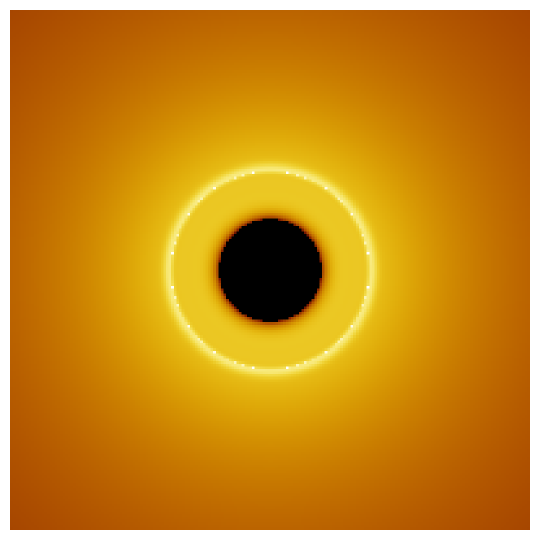}}
\subfigure[\tiny][$~q=0.39,~\theta_{obs}=17^{o}$]{\label{b105}\includegraphics[width=5.9cm,height=5.8cm]{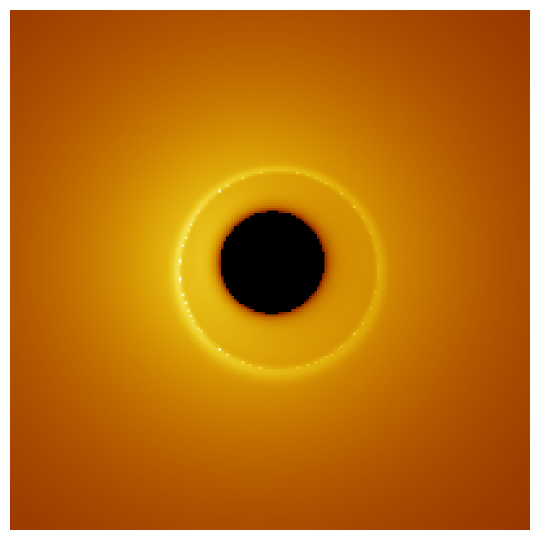}}
\subfigure[\tiny][$~q=0.39,~\theta_{obs}=80^{o}$]{\label{c106}\includegraphics[width=5.9cm,height=5.8cm]{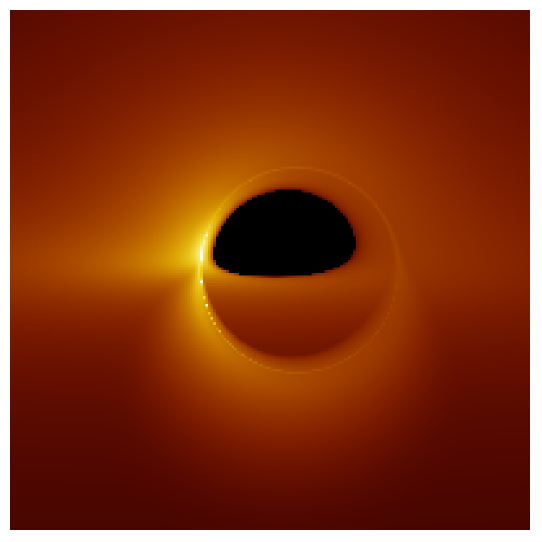}}
\subfigure[\tiny][$~q=0.69,~\theta_{obs}=0.01^{o}$]{\label{a107}\includegraphics[width=5.9cm,height=5.8cm]{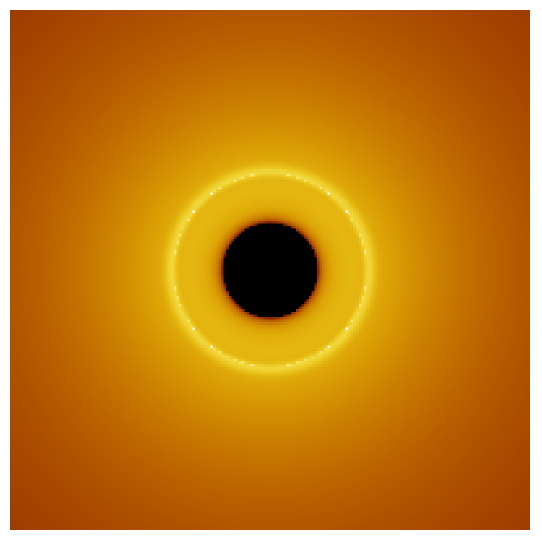}}
\subfigure[\tiny][$~q=0.69,~\theta_{obs}=17^{o}$]{\label{b108}\includegraphics[width=5.9cm,height=5.8cm]{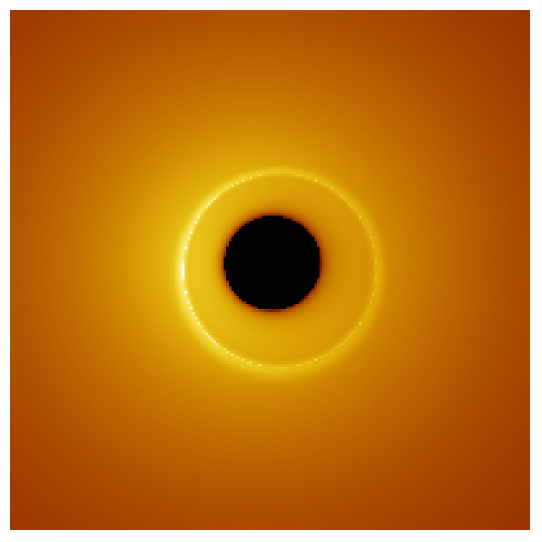}}
\subfigure[\tiny][$~q=0.69,~\theta_{obs}=80^{o}$]{\label{c109}\includegraphics[width=5.9cm,height=5.8cm]{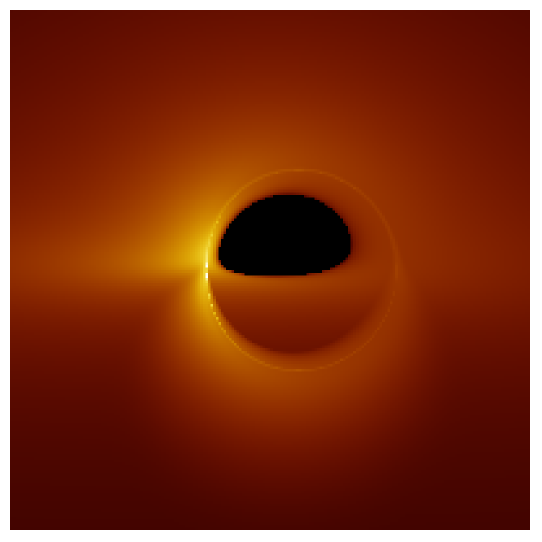}}
\caption{Density profiles of GK rotating BH illuminated by prograde
flows. Columns from left to right display models with inclination
angle $\theta_{obs}=0.01^{o},~17^{o},~80^{o}$ and rows from top to
bottom display with NED parameters $q=0.09,~0.39,~0.69$. The filled
black area is the BH event horizon. For all images, we set $a=0.69$,
$M=1$ and the observer's position is fixed at
$r_{obs}=500M$.}\label{shd101}
\end{center}
\end{figure*}

\begin{figure*}
\begin{center}
\subfigure[\tiny][$~q=0.001$]{\label{a101}\includegraphics[width=5.9cm,height=5.8cm]{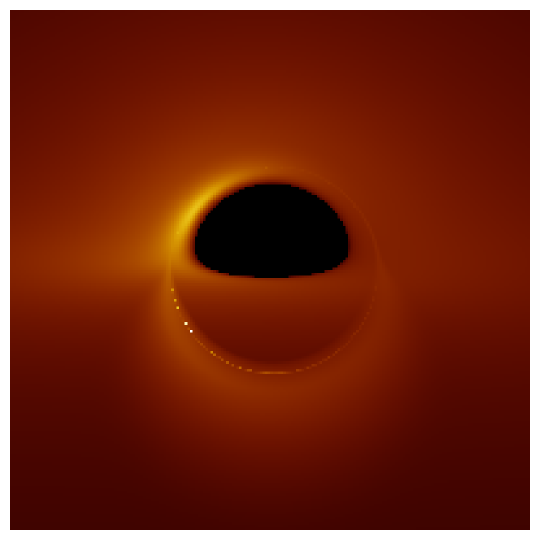}}
\subfigure[\tiny][$~q=0.01$]{\label{b102}\includegraphics[width=5.9cm,height=5.8cm]{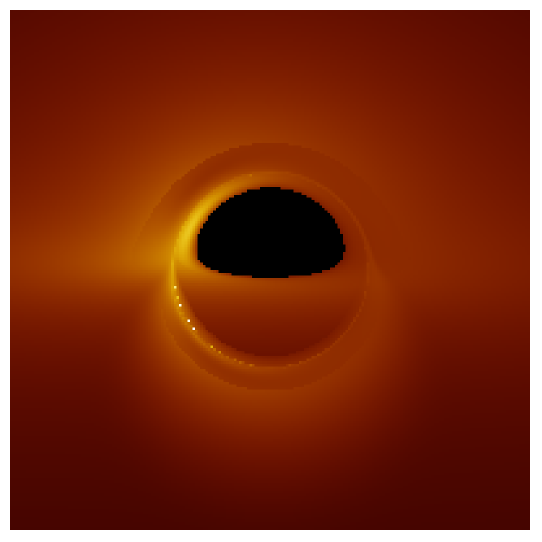}}
\subfigure[\tiny][$~q=0.99$]{\label{c103}\includegraphics[width=5.9cm,height=5.8cm]{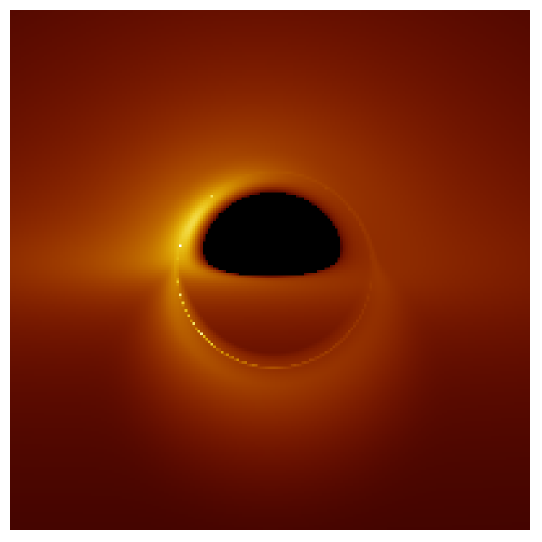}}
\caption{Density profiles of GK rotating BH is illuminated by prograde
flows. The filled black area is the BH’s event horizon. The white
dashed circle close to the lensed horizon known as ``photon ring''.
For all images, we fixed inclination angle $\theta_{obs}=~80^{o}$,
$a=0.69$, $M=1$ and the observer's position is fixed at
$r_{obs}=500M$.}\label{shd109}
\end{center}
\end{figure*}

\begin{figure*}
\begin{center}
\subfigure[\tiny][$~\theta_{obs}=0.01^{o}$]{\label{a110}\includegraphics[width=5.9cm,height=5.4cm]{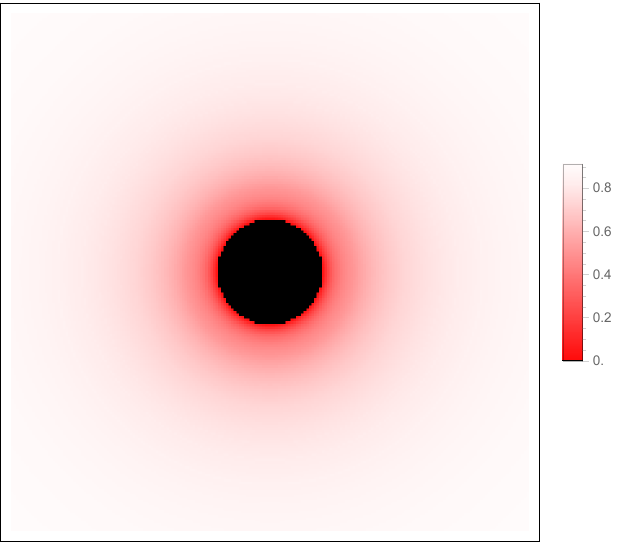}}
\subfigure[\tiny][$~\theta_{obs}=17^{o}$]{\label{b111}\includegraphics[width=5.9cm,height=5.4cm]{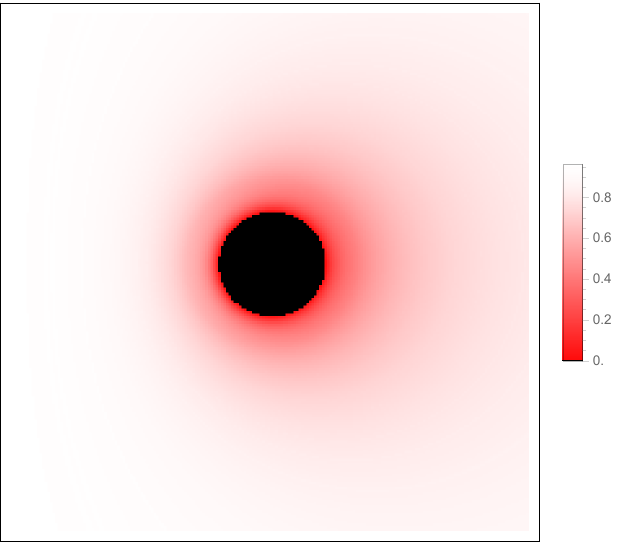}}
\subfigure[\tiny][$~\theta_{obs}=80^{o}$]{\label{c112}\includegraphics[width=5.9cm,height=5.4cm]{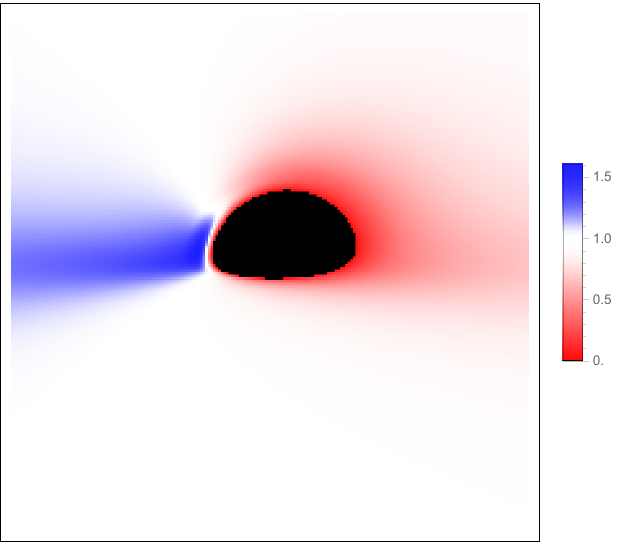}}
\subfigure[\tiny][$~q=0.09$]{\label{a113}\includegraphics[width=5.9cm,height=5.4cm]{Group12_2.pdf}}
\subfigure[\tiny][$~q=0.39$]{\label{b114}\includegraphics[width=5.9cm,height=5.4cm]{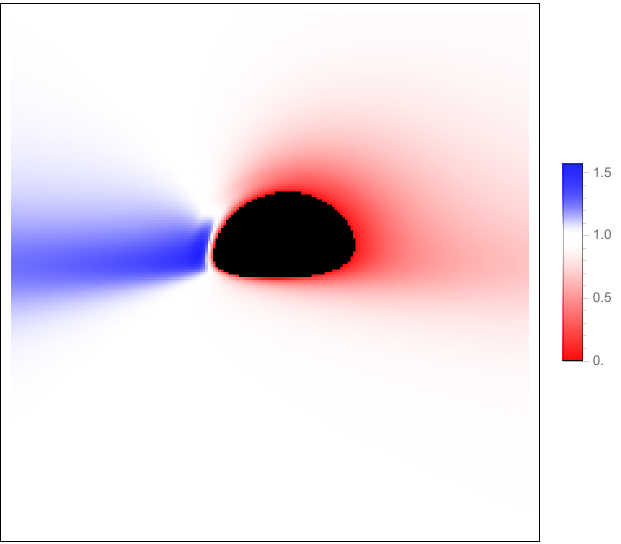}}
\subfigure[\tiny][$~q=0.69$]{\label{c115}\includegraphics[width=5.9cm,height=5.4cm]{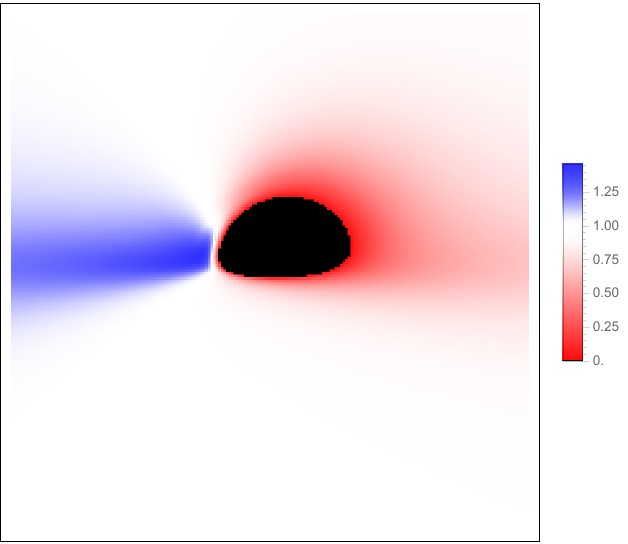}}
\subfigure[\tiny][$~a=0.09$]{\label{a116}\includegraphics[width=5.9cm,height=5.4cm]{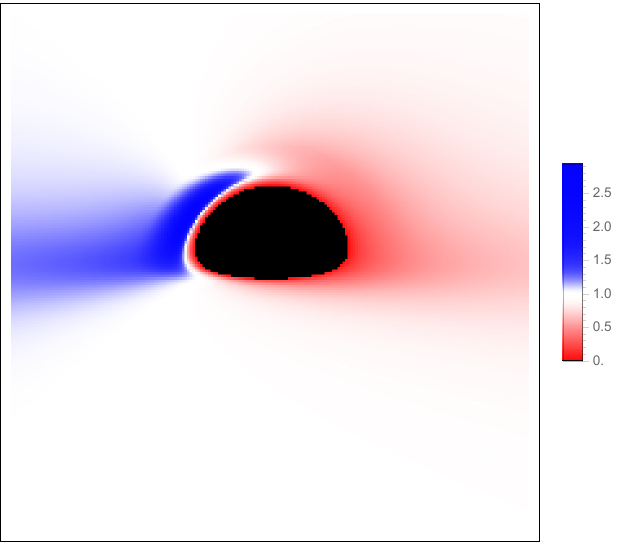}}
\subfigure[\tiny][$~a=0.69$]{\label{b117}\includegraphics[width=5.9cm,height=5.4cm]{Group12_2.pdf}}
\subfigure[\tiny][$~a=0.99$]{\label{c118}\includegraphics[width=5.9cm,height=5.4cm]{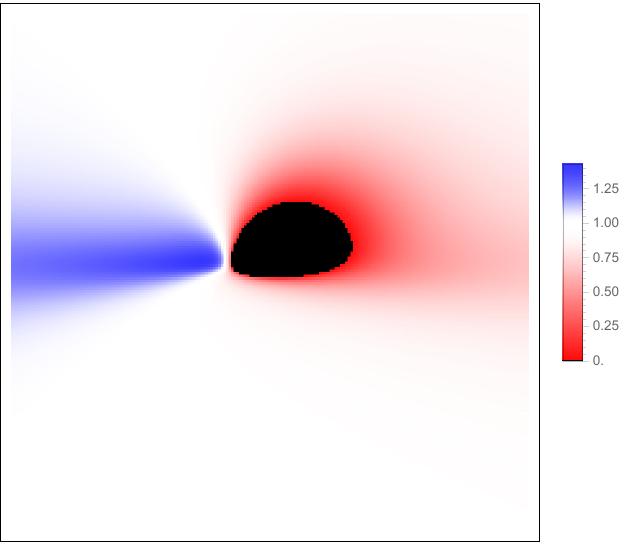}}
\caption{The direct images of the redshift factors of the
considering accretion disk model for different $\theta_{obs}$ with
fixed $a=0.69$.~$q=0.09$ (top row), for different $q$ with fixed
$a=0.69$.~$\theta_{obs}=80^{o}$, (middle row) and for different $a$
with fixed $q=0.09$.~$\theta_{obs}=80^{o}$, (bottom row). The red
and blue colors indicate the redshift and blueshift, respectively
and the filled black area is the BH’s event horizon. All these images
portray GK rotating BHs, illuminated by a prograde thin equatorial
accretion disk with $M=1$ and the observer's position is fixed at
$r_{obs}=500M$.}\label{shd102}
\end{center}
\end{figure*}

\begin{figure*}
\begin{center}
\subfigure[\tiny][$~\theta_{obs}=0.01^{o}$]{\label{a119}\includegraphics[width=5.9cm,height=5.4cm]{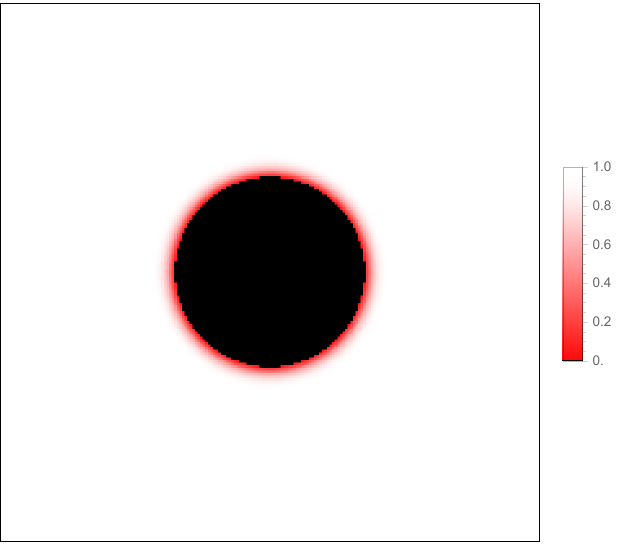}}
\subfigure[\tiny][$~\theta_{obs}=17^{o}$]{\label{b120}\includegraphics[width=5.9cm,height=5.4cm]{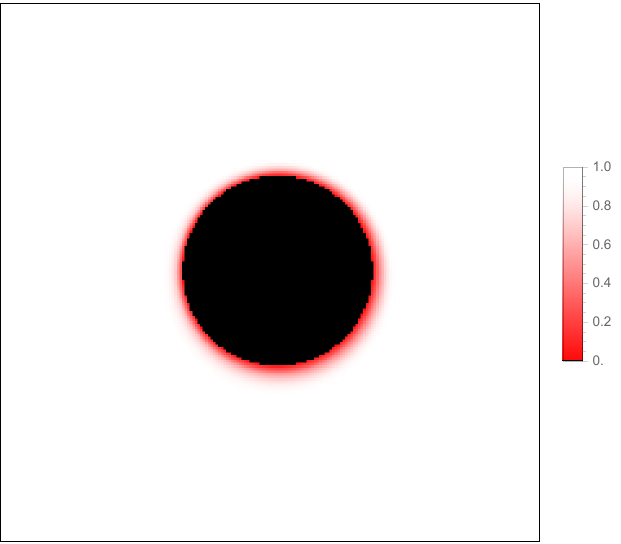}}
\subfigure[\tiny][$~\theta_{obs}=80^{o}$]{\label{c121}\includegraphics[width=5.9cm,height=5.4cm]{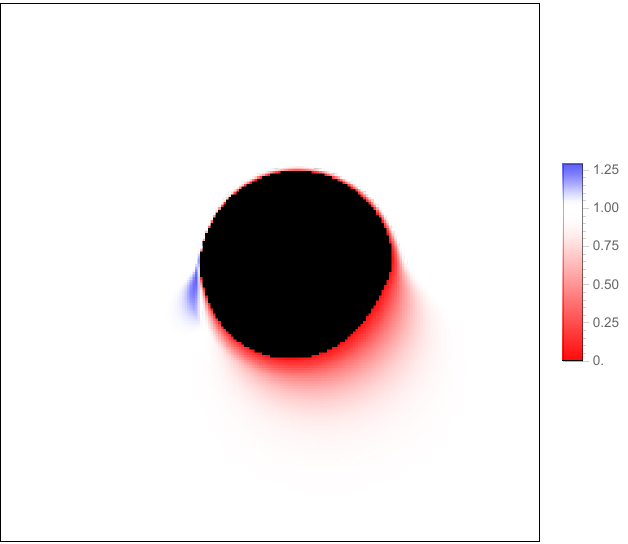}}
\subfigure[\tiny][$~q=0.09$]{\label{a122}\includegraphics[width=5.9cm,height=5.4cm]{Group12_3.pdf}}
\subfigure[\tiny][$~q=0.39$]{\label{b123}\includegraphics[width=5.9cm,height=5.4cm]{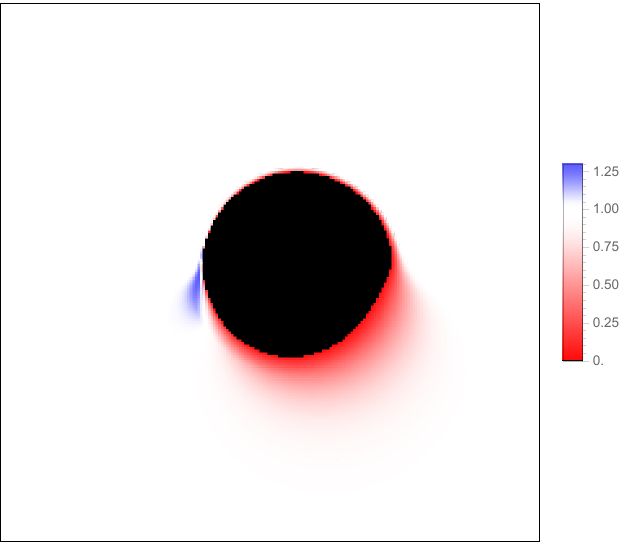}}
\subfigure[\tiny][$~q=0.69$]{\label{c124}\includegraphics[width=5.9cm,height=5.4cm]{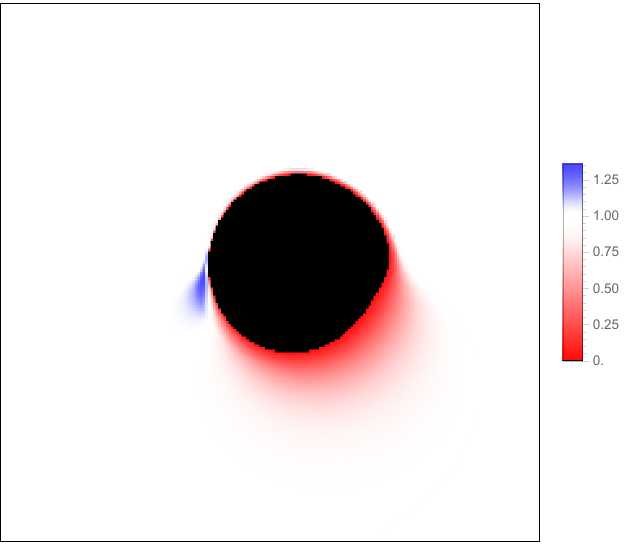}}
\subfigure[\tiny][$~a=0.09$]{\label{a126}\includegraphics[width=5.9cm,height=5.4cm]{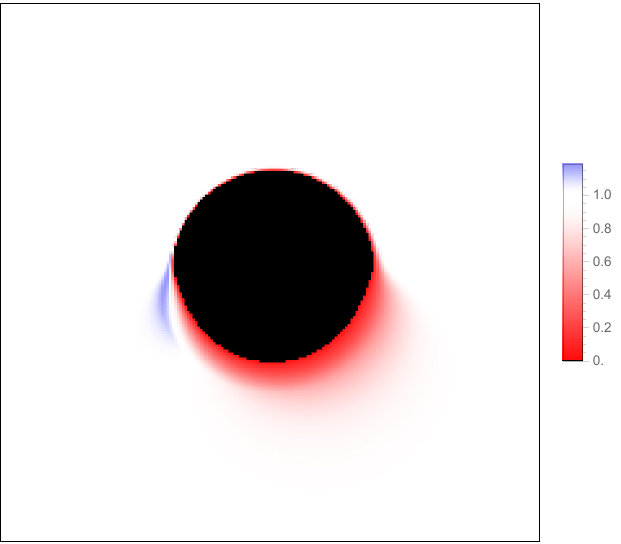}}
\subfigure[\tiny][$~a=0.69$]{\label{b127}\includegraphics[width=5.9cm,height=5.4cm]{Group12_3.pdf}}
\subfigure[\tiny][$~a=0.99$]{\label{c128}\includegraphics[width=5.9cm,height=5.4cm]{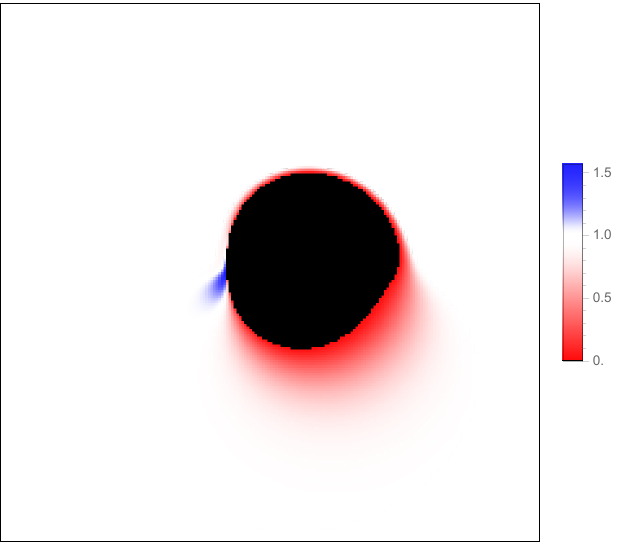}}
\caption{The lensed images of the redshift factors of the
considering accretion disk model for different $\theta_{obs}$ with
fixed $a=0.69$.~$q=0.09$ (top row), for different $q$ with fixed
$a=0.69$.~$\theta_{obs}=80^{o}$, (middle row) and for different $a$
with fixed $q=0.09$.~$\theta_{obs}=80^{o}$, (bottom row). The red
and blue colors indicate the redshift and blueshift, respectively
and the filled black area is the BH’s event horizon. All these images
portray GK rotating BHs illuminated by a prograde thin equatorial
accretion disk with $M=1$ and the observer's position is fixed at
$r_{obs}=500M$.}\label{shd103}
\end{center}
\end{figure*}

\begin{figure*}
\begin{center}
\subfigure[\tiny][$~q=0.001$]{\label{a110}\includegraphics[width=4.3cm,height=4cm]{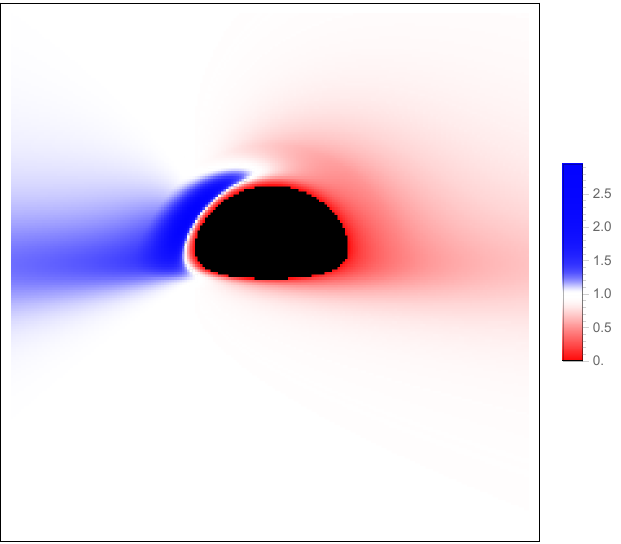}}
\subfigure[\tiny][$~q=0.005$]{\label{b111}\includegraphics[width=4.3cm,height=4cm]{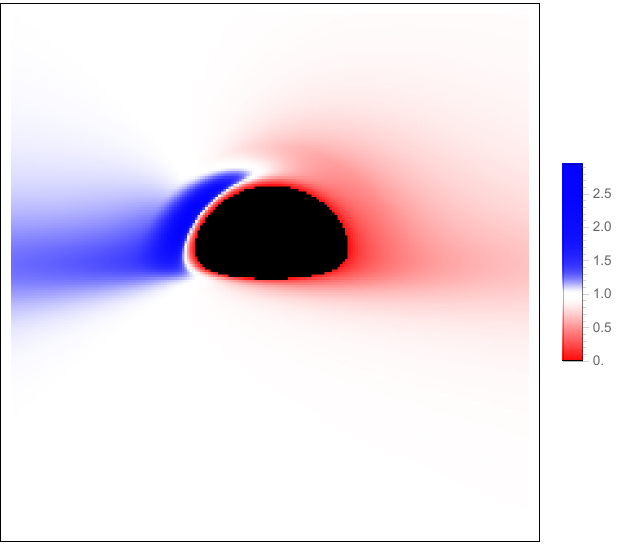}}
\subfigure[\tiny][$~q=0.01$]{\label{c112}\includegraphics[width=4.3cm,height=4cm]{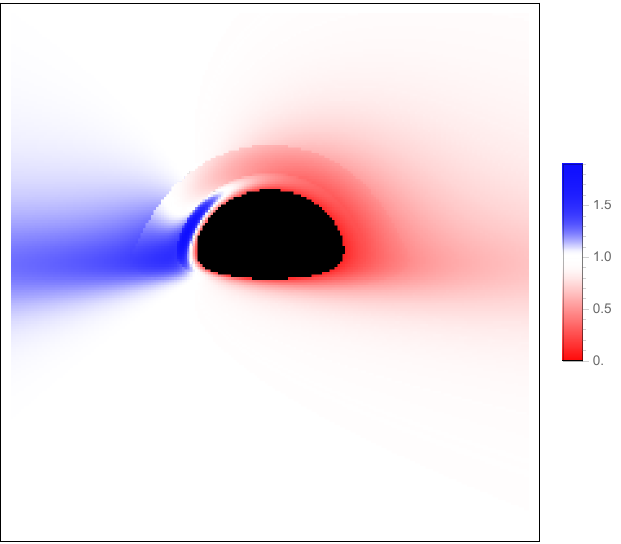}}
\subfigure[\tiny][$~q=0.015$]{\label{c112}\includegraphics[width=4.3cm,height=4cm]{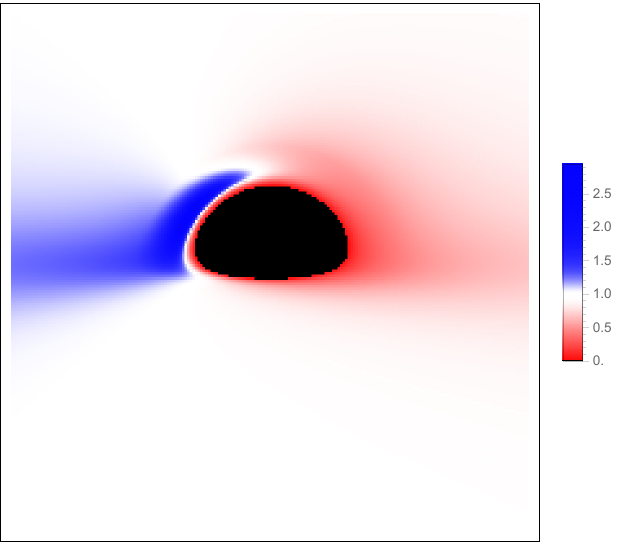}}
\caption{The direct images of the redshift factors of the
considering accretion disk model for different $q$ with inclination
angle $\theta_{obs}=~80^{o}$. The red and blue colors indicate the
redshift and blueshift, respectively and the filled black area is
the BH’s event horizon. All these images portray GK rotating BHs
illuminated by a prograde thin equatorial accretion disk with
$a=0.09$,~$M=1$ and the observer's position is fixed at
$r_{obs}=500M$.}\label{shd110}
\end{center}
\end{figure*}

In this subsection, we will briefly define the physical
interpretation of the prograde accretion disks surrounding the GK
rotating BH within our theoretical framework. In Fig.
\textbf{\ref{shd101}}, we illustrate the density profiles
illuminated by prograde flows for different values of $q$ and the
observational angle $\theta_{obs}$ with fixed values of $a=0.69$ and
$M=1$. Initially, in the left and middle columns of Fig.
\textbf{\ref{shd101}} where $\theta_{obs}=0.01^{o}$ and $17^{o}$,
respectively, the observational representation of bright band around
the BH shows that the intensity distribution is so sharp that the
direct and lensed images are hard to distinguish. However, with
increasing observational angle $\theta_{obs}=80^{o}$, as
shown in the right column of Fig. \textbf{\ref{shd101}}, from top to
the bottom one can see that the luminosity of the accretion disk is dim
as compared to the previous one. Moreover, a gradual metamorphosis
transpires appeared, wherein the disk-like structure gradually
transforms into a configuration reminiscent of a hat. And here, at
the event horizon $r=r_{+}$, the boundary of the hat-like shape of
the BH shows direct images of the accretion disk and the region
outside the white dashed ring (which is hard to see) shows the
lensed images of the accretion disk. In addition, the radiation flux
intensity illustrates a progressively growing asymmetry between the
left and right facets of the accretion disk. This emphasized
asymmetry is rooted in the heightened influence of the Doppler
effect with the addition of the observation angle. Consequently, the
disparity in radiant flux intensity between the left and right sides
of the accretion disk magnifies significantly with the aid of the
observation angle.

The origin of this phenomenon can be observed in the relative motion
between the observer and the accretion flows emitted within the
confines of the accretion disk. Note that, the inner shadow is from
the fact that a light ray cannot obtain any intensity if directly
falling into the horizon without passing the equatorial plane such
as $m=0$. In Fig. \textbf{\ref{shd109}}, we further present the
density profiles for three different values of NED parameter $q$
with fixed inclination angle $\theta_{obs}=~80^{o}$, $a=0.69$ and
$M=1$. Here we observe that initially, when $q$ has smaller values
such as $q=0.001$, the white dashed circle only shows at the bottom of
the hat-like shape and on the left side of the screen, we see
significant Doppler effects due to the forward rotation of the
prograde accretion disk. All these effects can be seen more clearly
when $q=0.01$ and $0.99$. This leads to changes in the
NED parameter $q$, each image exhibits a distinct photon ring as
well as Doppler effects on the left side of the screen.
\begin{center}
{\footnotesize{\bf Table 1.} The maximal blueshift under different
spin and NED parameters $a$ and $q$, respectively. All data are
sourced from direct images,
with $M=1$, an observational angle of $\theta_{obs}= 80^\circ$, and prograde motion.}\\
\vspace{2mm}
\begin{tabular}{c|c|c|c|c|c|c|c}
\hline
   \multicolumn{8}{c}{$\textbf{g}_{\text{max}} $ }  \\
\hline \diagbox{q}{a} & 0.09 & 0.19 & 0.29 & 0.39 & 0.49 & 0.59 &
0.69 \\ \hline 0.09 & 2.9021 & 2.4505 & 2.1595  & 1.9417 & 1.7992 &
1.6943 & 1.6065 \\ \hline 0.19 & 2.8948 & 2.4341 & 2.1437 & 1.9274 &
1.7882 & 1.6838 & 1.5967 \\ \hline 0.29 & 2.8379 & 2.4033 & 2.1152 &
1.9139 & 1.7686 & 1.6749 & 1.5791
\\ \hline 0.39 & 2.7731 & 2.3554 & 2.0723 & 1.8909 & 1.7425 &
1.6578 & 1.5616 \\ \hline 0.49 & 2.6823  & 2.2881 & 2.0220 & 1.8560
& 1.7168 & 1.6298 & 1.5436 \\ \hline 0.59 & 2.5654 & 2.2008 & 1.9835
& 1.8072 & 1.6836 & 1.5941 & 1.5084 \\ \hline 0.69 &
2.4450 & 2.1433 & 1.9252 & 1.7696 & 1.6483 & 1.5580 & 1.4607 \\
\hline
\end{tabular}
\end{center}

\begin{center}
{\footnotesize{\bf Table 2.} The maximal blueshift under different
spin and NED parameters $a$ and $q$, respectively. All data are
sourced from lensed images,
with $M=1$, an observational angle of $\theta_{obs}= 80^\circ$, and prograde motion.}\\
\vspace{2mm}
\begin{tabular}{c|c|c|c|c|c|c|c}
\hline
   \multicolumn{8}{c}{$g_{\text{max}} $ }     \\
\hline \diagbox{q}{a} & 0.09 & 0.19 & 0.29 & 0.39 & 0.49 & 0.59 &
0.69 \\ \hline 0.09 & 1.1873 & 1.1955 & 1.2028 & 1.2161 & 1.2345 &
1.2529 & 1.2694 \\ \hline 0.19 & 1.1712 & 1.1899 & 1.2032 & 1.2160 &
1.2327 & 1.2499 & 1.2693 \\ \hline 0.29 & 1.1786 & 1.1886 & 1.2031 &
1.2172 & 1.2328 & 1.2507 & 1.2662
\\ \hline 0.39 & 1.1855 & 1.1889 & 1.2017 & 1.2169 & 1.2327 &
1.2483 & 1.2565 \\ \hline 0.49 & 1.1900 & 1.1942 & 1.2002 & 1.2136 &
1.2283 & 1.2472 & 1.3094 \\ \hline 0.59 & 1.1939 & 1.2037 & 1.2152 &
1.2284 & 1.2533 & 1.2875 & 1.3167 \\ \hline 0.69 & 1.1967 & 1.2098 &
1.2235 & 1.2422 & 1.2681 & 1.2897 & 1.3275
\\ \hline
\end{tabular}
\end{center}

\section{Examining the Propagation of Redshift Factors}

Since, the relative motion between the accretion disk and the
distant observer, provides a deeper understanding of the intricate
interplay among BHs and the resulting properties of accretion disks.
These significant results can be understood with the help of the
Doppler’s effect. For this, in the imaging process of the BH, we
demonstrate an accurate assessment of the redshift factors
associated with the behavior of emitting particles.

In Figs. \textbf{\ref{shd102}} and \textbf{\ref{shd103}}, we present
the redshifts factors for different values of model parameters.
These include the direct images ($m=1$) and lensed images ($m=2$) of
the prograde accretion disk. When $\theta_{obs}$ has smaller values
as shown in top rows of Figs. \textbf{\ref{shd102}} and
\textbf{\ref{shd103}}, no significant influence of the
blueshift on the observer's screen is observed. However, when
$\theta_{obs}=80^{o}$, a significant appearance of the
blueshift has been observed on the observer's screen. This phenomenon will happen due
to the observer location, the dynamics of the accretion disk
rotation, and the significant effect of gravitational redshift from
the BHs strong gravitational field. In the second and third rows of
Figs. \textbf{\ref{shd102}} and \textbf{\ref{shd103}}, we present
The accretion phenomena with fixed position of the observer such as
$\theta_{obs}=80^{o}$ and analyze the nature of other involved parameters.
Here, in the case of direct images, the observer sees a clear
bright optical appearance of redshift and blueshift factors on the
screen. The variation of parameters corresponds to changes in
brightness, indicating different temperatures and densities in the
disk. In the case of lensed images, due to the formation of some
extra rings or arcs around the BH, the observer can see more prominent redshift
factor as compared to blueshift on the screen, see the
second and the third rows of Fig. \textbf{\ref{shd103}}. In Fig.
\textbf{\ref{shd110}}, we present the influence of smaller values
of NED parameter $q$ on the direct images of accretion disk.
Initially, when $q$ has smaller values, both red and blue shifts are
Negligible, merge with each other. However, with the aid of $q$
such as, $q=0.015$ the redshift moves to the left side of the screen and
slightly merge with blueshift. In all these cases, the redshift is
closely lie to ISCO. In Tables. {\bf1} and {\bf2}, we present the
$\textbf{g}_{\text{max}}$ of direct and lensed images, respectively
under various values of spin and NED parameters. It is observed
from Table. {\bf1} that $a$ and $q$ help to decrease the
strength of maximal blueshift factor of direct images on the screen.
However, in the case of lensed images, as presented in Table. {\bf2},
the maximal strength of blueshift factor on the screen is increased
with respect to both spin and NED parameters. All these results
are clearly reflects on both direct and lensed images on the screen
for prograde accretion disk.

\begin{figure*}
\begin{center}
\subfigure[\tiny][$~q=0.09$]{\label{a129}\includegraphics[width=5.9cm,height=5.4cm]{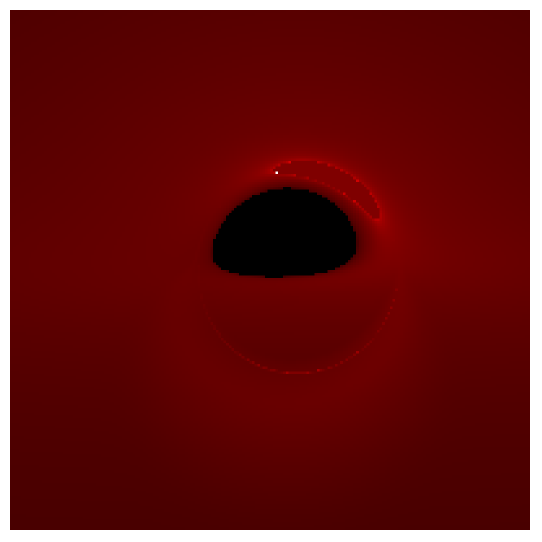}}
\subfigure[\tiny][$~q=0.39$]{\label{b130}\includegraphics[width=5.9cm,height=5.4cm]{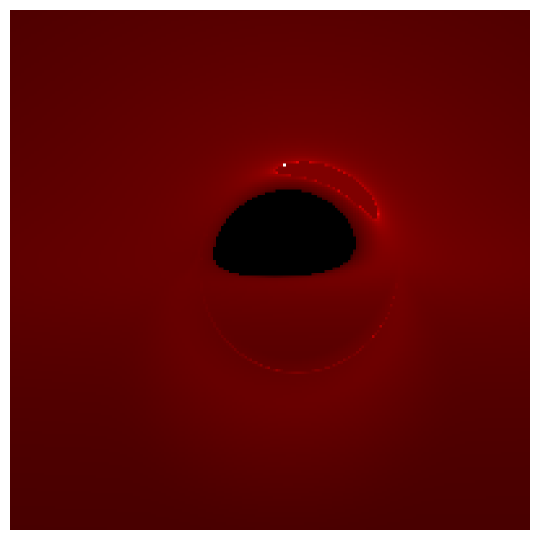}}
\subfigure[\tiny][$~q=0.69$]{\label{c131}\includegraphics[width=5.9cm,height=5.4cm]{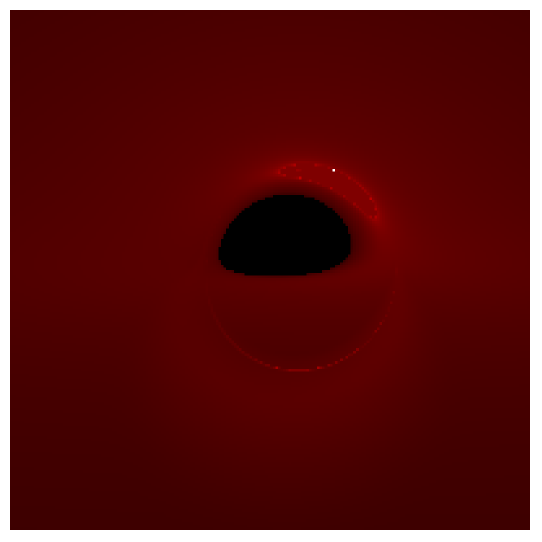}}
\caption{Density profiles of GK rotating BH are illuminated by
retrograde flows. The filled black area is the BH’s event horizon. The
white dashed circle close to the lensed horizon known as ``photon
ring'', which is hard to see here. The slightly prominent white
dashed curve is the lensed images of the accretion disk. For all
images, we fixed inclination angle $\theta_{obs}=~80^{o}$, $a=0.69$,
$M=1$ and the observer's position is fixed at
$r_{obs}=500M$.}\label{shd108}
\end{center}
\end{figure*}

Now, we are going to describe the geometric details of the
retrograde accretion disk observed on the observer's screen. As
depicted in Fig. \textbf{\ref{shd108}}, the observational
representation is suppressed by the gravitational redshift in the
scenario of retrograde accretion disk flow. At that frequency, the
optical depth suppressing the appearance of the lensed subring,
showing on the upper north half of the image on the screen. And the
optical depth is more suppressed with the aid of NED parameter $q$
and the optical appearance of the inner shadow is still suppressed.
This is due to, during the simulation, the emitting material
contributing to the increased total optical depth is contained
within the equatorial plane, and the jet material in front of the
event horizon shows low brightness. As a result, the direct
geometrical effect of the equatorial emission being truncated at the
event horizon is still visible, even it is hardly to see
\cite{rtb32}. These observations distinctly underscore the profound
impact of the NED parameter $q$ on the luminosity and structural
stability attributes of the accretion disk that encircles the BH.

\begin{figure*}
\begin{center}
\subfigure[\tiny][$~q=0.09$]{\label{a129}\includegraphics[width=5.9cm,height=5.4cm]{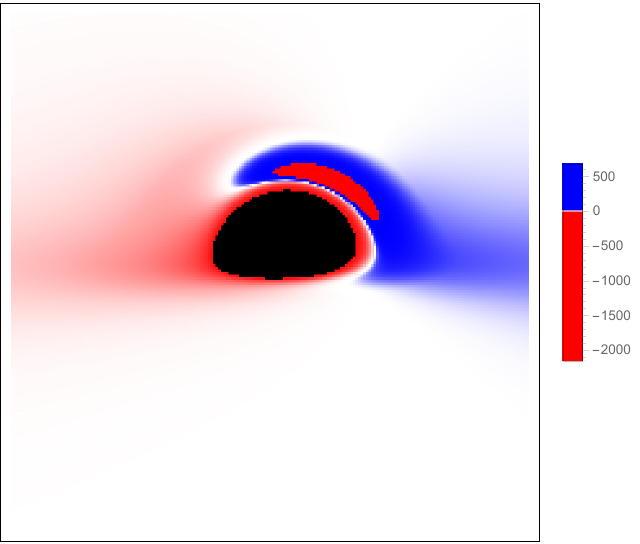}}
\subfigure[\tiny][$~q=0.39$]{\label{b130}\includegraphics[width=5.9cm,height=5.4cm]{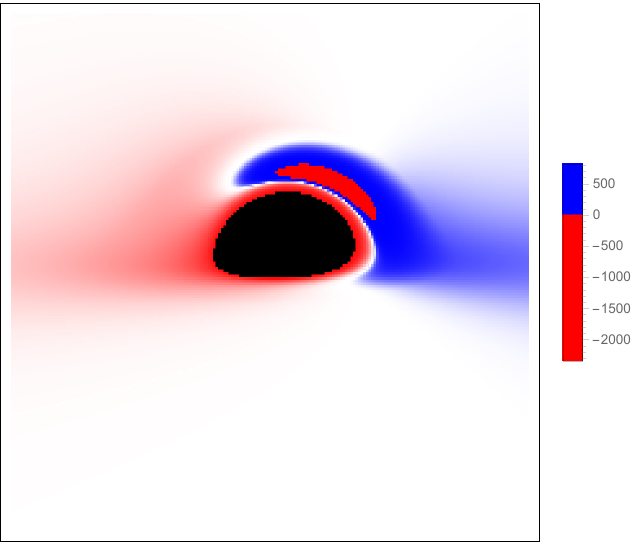}}
\subfigure[\tiny][$~q=0.69$]{\label{c131}\includegraphics[width=5.9cm,height=5.4cm]{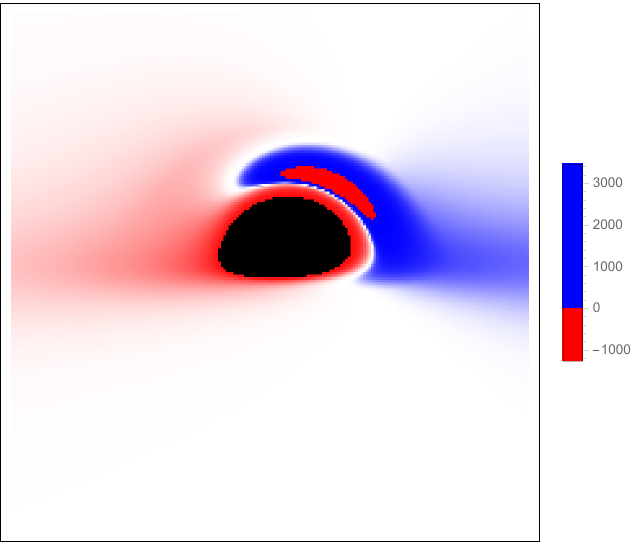}}
\subfigure[\tiny][$~q=0.09$]{\label{a135}\includegraphics[width=5.9cm,height=5.4cm]{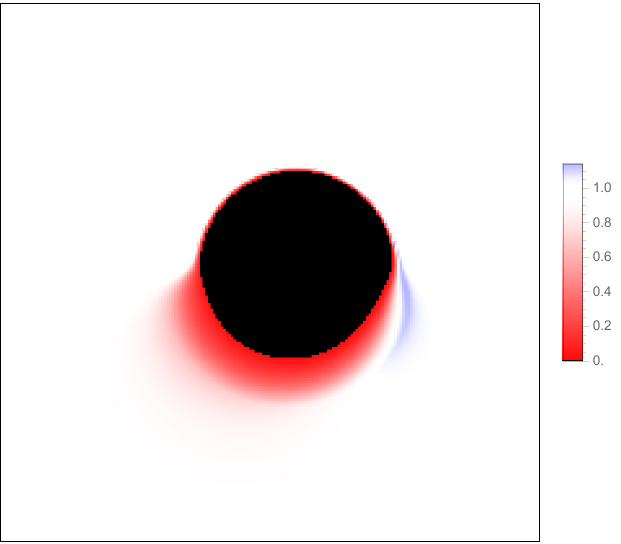}}
\subfigure[\tiny][$~q=0.39$]{\label{b136}\includegraphics[width=5.9cm,height=5.4cm]{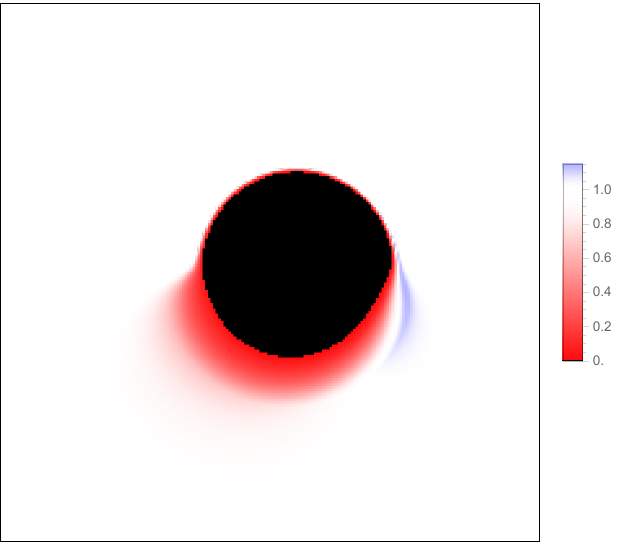}}
\subfigure[\tiny][$~q=0.69$]{\label{c137}\includegraphics[width=5.9cm,height=5.4cm]{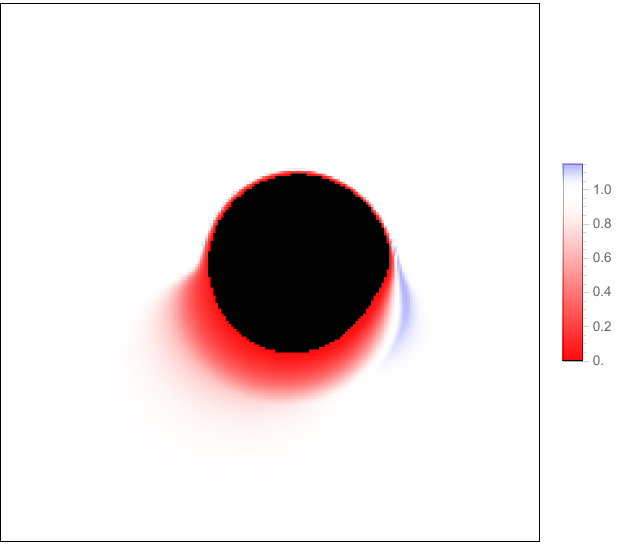}}
\caption{The direct and lensed images of the redshift factors of the
considering accretion disk model for different $q$ are illustrated
in top and bottom rows, respectively. The red and blue colors
indicate the redshift and blueshift, respectively and the filled
black area is the BH’s event horizon. All these images portray GK
rotating BHs that are illuminated by a retrograde thin equatorial accretion
disk with $a=0.69$, $\theta_{obs}=80^{o}$, $M=1$ and the observer's
position is fixed at $r_{obs}=500M$.}\label{shd111}
\end{center}
\end{figure*}

\begin{figure*}
\begin{center}
\subfigure[\tiny][$~\theta_{obs}=0.01^{o}$]{\label{a129}\includegraphics[width=5.9cm,height=5.4cm]{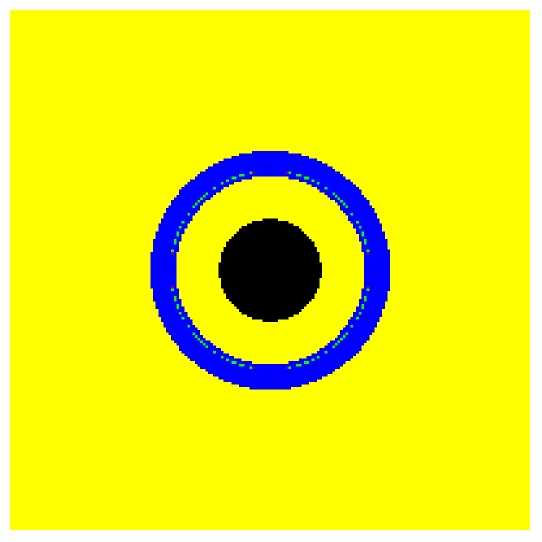}}
\subfigure[\tiny][$~\theta_{obs}=17^{o}$]{\label{b130}\includegraphics[width=5.9cm,height=5.4cm]{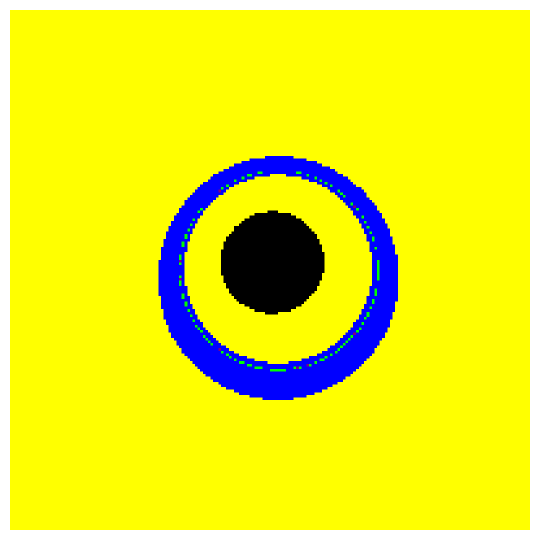}}
\subfigure[\tiny][$~\theta_{obs}=80^{o}$]{\label{c131}\includegraphics[width=5.9cm,height=5.4cm]{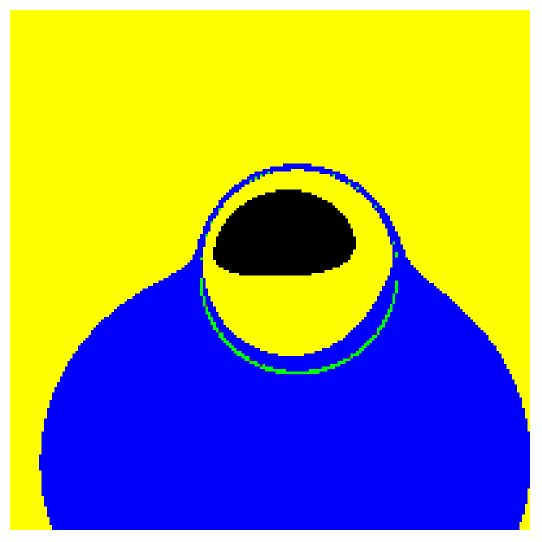}}
\caption{The optical illustrations of direct, lensed and photons
ring emissions for different inclination angle $\theta_{obs}$. The
yellow, blue and the green band colors correspond to the direct,
lensed and photon ring emissions, respectively. All these images
portray GK rotating BHs illuminated by a retrograde thin equatorial
accretion disk with $a=0.69$, $q=0.39$, $M=1$ and the observer's
position is fixed at $r_{obs}=500M$.}\label{shd113}
\end{center}
\end{figure*}

In Fig. \textbf{\ref{shd111}}, we presented the redshift factors for
direct and lensed images in the top and bottom rows under different
values of $q$. And the accretion flow is retrograded in both cases.
In the first row, we notice that the redshift near the ISCO, which
will be suppressed by an increase of $q$. And in the right
side of the screen, the influence of redshift lensed curves also
appears in the emission of blue shift. This will happen due to the
retrograde accretion flow and relativistic jet which lies in front
of the bulk of emission in the equatorial plane. In the bottom
panel, only on the right side of the screen, a small petal of the
blueshift appeared and the redshift factor also lie close to the
ISCO. The luminosity of the accretion disk is significantly
decreased, because the contribution of the emission flow is smaller
due to the lensed images phenomena. consequently, in these
simulations, the emission from accretion disk are retrograde
emission, which could produce a central brightness depression with a
slightly smaller area than the equatorial inner shadow, as the
emission region intersects with the event horizon at a maximum
latitude on the equatorial plane. We further exhibit the non-uniform
resolution used for image computation and storage in Fig.
\textbf{\ref{shd113}}. Here the resolution of the bands contains
roughly the same number of pixels, resulting in a more manageable
number of pixels per image. The complete null geodesic of the GK
rotating spacetime is presented with three different bands such as
yellow, blue and green corresponding to the rays that cross the
equatorial plane of the BH once, twice, and three times,
respectively. In all these cases, the photon rings always lie
exactly within the blue and yellow lensing bands. However, the
confine shape of blue bands is deformed with change in the
observational position and reach the maximum when
$\theta_{obs}=~80^{o}$. Hence, the overall discussion about the
observation of a BH's gravitational lensing effect provides an
avenue for constraining the involved model parameters on the scale
of BHs. We hope these observations may serve as a cornerstone for
refining our comprehension of GK rotating BH and its behavior on
astronomical scales.

\section{Conclusions and Discussion}

In the last few years, the study of BH spacetime geometry with light
rings and accretion disk has become increasingly popular
among scientific community, particularly after the beginning of
multi-messenger astronomy. The existence of accretion disks
surrounding BHs is a well-substantiated phenomenon prompted by a
wealth of empirical evidence. An important milestone was reached
with the accomplishments of the EHT, which provided some deep
insights into BH spacetime geometry. Particularly, the iconic
depiction of a super massive BH within the M$87$ galaxy captured a
photon ring of luminosity close to the event horizon, surrounded by
an accretion disk. These observations revealed that the presence of
a bright ring of radiation enclosing a central brightness
depression provides the intrinsic properties of the plasma
radiation near the BH. In this regard, the imaging process of BHs,
the accretion of matter surrounded by high-energy radiation plays a
significant role. To exhibit the results on the observer's screen,
we considered different background light sources, i.e., the
spherical background light source and the thin disk accretion model.
Once the simulated images are displayed on the observer's screen, we
have identified the influence of the observable properties of the
celestial light source and the thin accretion disks stemming from
the changes in the spacetime structure due to variation of
parameters within the fabric of GK rotating BH. Particularly, this
study has unveiled profound connections between parameters
$a,~q$, $~\theta_{obs}$ and the optical appearance of the shadow
images of spherical background light source as well as the accretion
disk.

To analyze the influence of spacetime parameters, we
investigated the shadow images for a nearer observer as depicted in
Fig. \textbf{\ref{shd1}}. It is noticed that spin moves the
shadow to the rightward of the screen and possible flatness is
observed when $a=1$, see Fig. \textbf{\ref{shd1}} (a). On the other
hand, when we varied the NED parameter $q$, the shadow radius
decreased with the help of $q$ as shown in Fig.
\textbf{\ref{shd1}} (b). Both parameters have the
significant influence on the shadow radius as expected, which we
also observed in latter discussion. By considering an observer with zero angular
momentum within the domain of outer communication, the
precise shape of the shadow projected on the observer screen can be
obtained with the help of backward ray-tracing technique in
celestial coordinates. The corresponding results interpret that the
observational plane of the observer having central
dark region, and outside the ``D'' shape petals, there is a white dashed
circle. The filled dark region represents the BH shadow and white
dashed circle directly demonstrated the Einstein ring, see Figs.
\textbf{\ref{shd2}}-\textbf{\ref{shd4}}.

Initially, we observe that NED parameter $q$ is responsible for slightly increase
the radius of Einstein ring but the size of the
BH shadow is constant. Moreover, at $q=1$, the resulting Einstein
rings are gradually started to transform into arc-like shapes around
the D-shaped configuration. When we further increase the influence
of spin parameter $a$ and vary $q$ from left to right, the size of
the BH shadow is slightly increased, and the Einstein ring shows a
transition from an axisymmetric closed circle to an arc-like shape
on the left and right side of the screen. These observations clearly
indicate that the increasing values of $a$ and $q$
led to an enhancement in the degree of deformation of the shadow
shape, and the optical appearance of shadow images also undergoes
distortion due to the drag effect. The aforementioned
details reflect the description of the Figs.
\textbf{\ref{shd2}}-\textbf{\ref{shd4}}, where we fixed the
observation angle $\theta_{obs}=80^{o}$. Next, we imposed a
straightforward approach of spherical background light source model
and analyzed the geometric details of BH images with thin accretion
disk model. In the real world scenario, most of the plasma rotating
around the BHs spacetime that undergoes gravitational effects and
radiate high-energy particles, thereby giving rise to luminous
accretion disks around the BH. Based on backward ray-tracing method,
we further explored the dynamics of bright accretion disk as the
fundamental source of background light source and the under consideration
accretion model is assumed as an optically and geometrically thin
accretion disk.

In Fig. \textbf{\ref{shd101}}, we plot the BH shadow images of thin
accretion disk model for different values of $q$ and $\theta_{obs}$,
while keeping the spin parameter $a$ is constant. In all these
images, one can always notice a filled dark region surrounded by a
narrow photon ring. The dark region corresponds to accretion disk
profile at $r=r_{+}$, so called the inner shadow while the photon
ring closely aligned with the critical curve of the BH. In the first
column of Fig. \textbf{\ref{shd101}}, when $\theta_{obs}=0.01^{o}$,
the dark region always lies in the center of the screen and there is
a very dim strip of concentric circles appeared with the
luminous ring. This will happen due to the sight line which is
perpendicular to the disk and the accreting particles of the disk
has no component at that sight line. This phenomenon is resulting in a
little influence of Doppler effect in the redshift factor along
with only gravitational redshift. When we increase the inclination
angle $\theta_{obs}=17^{o}$, as shown in the middle column of Fig.
\textbf{\ref{shd101}}, one can notice that the gradual deformation
in the dark central region, slightly moving towards the right
bottom of the screen and there is no significant exhibition of
concentric circle with the photon ring. When we further increase the
value of inclination angle $\theta_{obs}=80^{o}$, the deformation
can be clearly identified where the central dark region transforms
into a hat-like shape, see the right column of Fig.
\textbf{\ref{shd101}}. Moreover, a prominent feature also appeared
on the left side of the screen, where a luminous crescent or
eyebrow-shaped region appeared which can participate to
magnifies the angle, resulting in an increasingly pronounced Doppler
effect. In all these cases from top to bottom, distinct
features can be seen more clearly with the increase of NED
parameter $q$. Interestingly, when $\theta_{obs}$ has smaller
values, there is no significant separation between direct and lens
images, because the distribution of the observed luminosity is not
sharp enough. Hence, the change of observational angle does not
affect the location of the critical curve, but its emission profile
is slightly deformed with values of $\theta_{obs}$. In
addition, one can see the influence of NED parameter $q$ on shadow
images when we fixed $\theta_{obs}=80^{o}$, see Fig.
\textbf{\ref{shd109}}. Here, at smaller values of $q$, the
observational luminosity of Doppler effect is relatively smaller on
the left side of the screen as compared to larger one. This shows
that an increase in $q$ yields a stronger influence on the
luminosity of accretion matter.

In order to analyze the realistic scenario of accreting matter
around the BH, we further investigate the redshift factors of both
the direct and lensed images originating from the accretion disk.
Figures. \textbf{\ref{shd102}} and \textbf{\ref{shd103}}
illustrate redshifts factors for both direct and lensed images,
respectively under different values of model parameters.
Particularly, the top row of Fig. \textbf{\ref{shd102}} shows that
at smaller inclination angle $\theta_{obs}$, the observer only
see the redshift factors on the screen. However, with increasing
$\theta_{obs}$, one can also see the blueshift on the left
side of the screen. For the middle and the bottom rows of Fig.
\textbf{\ref{shd102}}, we fixed the inclination angle to
$\theta_{obs}=80^{o}$ and seen the influence of both NED and spin
parameters on the redshift factors. In all these images, the
observer can see both redshift and blueshift factors on the screen,
but the formation of a strict red ring structure close to the inner
shadow, which is attributed to the emission of light by particles
within the critical plunging orbits. In Fig. \textbf{\ref{shd103}},
we demonstrated the redshift factors of the lensed images, where the
parameter values correspond to those in Fig. \textbf{\ref{shd102}}.
Initially, at smaller values of $\theta_{obs}$, there is just a
prominent red ring structure surrounding the inner shadow. However,
when we increase $\theta_{obs}=80^{o}$, the red color is
visually represented by a continuous linear color map and on the
left side of the screen, there is only a petal like map of the blue
color, see the top row of Fig. \textbf{\ref{shd103}}. In the middle
and the bottom rows of Fig. \textbf{\ref{shd103}}, we fixed the value
of $\theta_{obs}=80^{o}$ and varied the relevant parameters such as
$q$ and $a$. Here, the similar observational appearance of the
redshift factors are observed as discussed in the previous case. All
these results showed that the size of the inner shadow region is
correlated with the values of the associated parameters, while its
metamorphosis is persuaded by the observation inclination. In Fig.
\textbf{\ref{shd110}}, we also plotted the direct images of the
redshift factors for smaller values of $q$ with inclination angle
$\theta_{obs}=80^{o}$ as an example. From these images, one can
observe that, in addition to the influence of the magnetic field on
the size of the image, there are obvious blueshift near the ISCO,
and both red and blue colors are merged with each other prominently.

In Fig. \textbf{\ref{shd108}}, density profiles of BH shadow images
illuminated by retrograde flows for different values of $q$. Here
the optical appearance is suppressed with enhancing
magnetic influence and morphology of the photon ring remains
unaffected in all these cases. A distinct bright area resembles a
crescent or an eyebrow emerges on the top right side of the screen,
where both gravitational redshift of the accretion disk are merged
due to the retrograde flows. We further illustrated an investigation
into the redshift of both the direct and lensed images originating
from the retrograde flows in Fig. \textbf{\ref{shd111}}. It can be
observed that the redshift or blueshift effect of both direct and
lens images is more obvious when the parameter $q$ has
larger values. A significant difference is observed here in the case of
lensed images, the effect of blueshift has appeared on the right side
of the screen, which is due to retrograde flows of accretion disk,
as shown in bottom panel of Fig. \textbf{\ref{shd111}}. For better
understanding about the direct and lensed images of the thin
accretion disk, we interpreted their observed fluxes under inclination
angle $\theta_{obs}$ in Fig. \textbf{\ref{shd113}}. From these
images, it can be noticed that the photon ring always lies within the
range of lensed image when $\theta_{obs}=0.01^{o}$, and showed the
small amount of observed flux of the lensed emission inside the
photon ring. When we increase the observed inclination angle such as
$\theta_{obs}=17^{o}$, both direct and lensed ring structures slightly
deformed, and the observed flux move towards the
lower half of the screen. And when $\theta_{obs}=80^{o}$, the
deformation of direct and lensed images is higher as compared to
smaller values of $\theta_{obs}$. Moreover, the majority of the
observed flux of lensed emission is concentrated in the bottom half
region of the screen and the central dark region is transformed into
hat-like metamorphosis.

Based on our present analysis, we conclude that the above discussion
may provide fruitful results for the theoretical study of the BH
shadows for direct and lensed images and their relevant dynamics.
Since rotating objects are considered to be more realistic in the
universe, we hope that these groundbreaking results inspire the
theoretical study of BH shadow images and other physical quantities
that may be helpful for the observational teams working on achieving
high resolution images of accretion disk.

\end{document}